\documentclass{emulateapj}
\usepackage{graphicx}
\usepackage{amsmath, amssymb, amsfonts}
\usepackage{natbib}
\newcommand{\xmm}{{\sl XMM-Newton}\xspace}

\newcommand{\esas}{{\sl XMM-ESAS}\xspace}
\newcommand{\sas}{{\sl SAS}\xspace}

\usepackage{xspace}

\shortauthors{Paterno-Mahler et al.}
\shorttitle{The Triple System A98}
\submitted{Submitted to the Astrophysical Journal, 2014/03/08}
\accepted{2014/07/10}
\begin{document}

\title{Merger Signatures in the Galaxy Cluster Abell~98}
\author{R. Paterno-Mahler\altaffilmark{1,5}, S.W. Randall\altaffilmark{2},  E. Bulbul\altaffilmark{2}, F. Andrade-Santos\altaffilmark{2}, E. L. Blanton\altaffilmark{1}, C. Jones\altaffilmark{2}, S. Murray\altaffilmark{2, 3}, and R.E. Johnson\altaffilmark{4}}
\altaffiltext{1}{Astronomy Department and Institute for Astrophysical Research, Boston University, 725 Commonwealth Ave, Boston, MA 02215, USA; rachelpm@bu.edu, eblanton@bu.edu}
\altaffiltext{2}{Harvard Smithsonian Center for Astrophysics, 60 Garden Street, Cambridge, MA 02138, USA; srandall@head.cfa.harvard.edu, ebulbul@head.cfa.harvard.edu, fsantos@cfa.harvard.edu, cjones@cfa.harvard.edu}
\altaffiltext{3}{Physics and Astronomy Department, Johns Hopkins University, 3400 N. Charles Street, Baltimore, MD, 21218, USA; ssm@pha.jhu.edu}
\altaffiltext{4}{Department of Physics, Gettysburg College, Gettysburg, PA 17325, USA; rjohnson@gettysburg.edu}
\altaffiltext{5}{Visiting student at Harvard Smithsonian Center for Astrophysics (2013/05-2013/08)}

\begin{abstract}
We present results from \textit{Chandra} and \xmm observations of Abell~98 (A98), a galaxy cluster with three major components: a relatively bright subcluster to the north (A98N), a disturbed subcluster to the south (A98S), and a fainter subcluster to the far south (A98SS).  We find evidence for surface brightness and temperature asymmetries in A98N consistent with a shock-heated region to the south, which could be created by an early stage merger between A98N and A98S.  Deeper observations are required to confirm this result.  We also find that A98S has an asymmetric core temperature structure, likely due to a separate ongoing merger.  Evidence for this is also seen in optical data.  A98S hosts a wide-angle tail (WAT) radio source powered by a central active galactic nucleus (AGN).  We find evidence for a cavity in the intracluster medium (ICM) that has been evacuated by one of the radio lobes, suggesting that AGN feedback is operating in this system.  Examples of cavities in non-cool core clusters are relatively rare.  The three subclusters lie along a line in projection, suggesting the presence of a large-scale filament.  We observe emission along the filament between A98N and A98S, and a surface brightness profile shows emission consistent with the overlap of the subcluster extended gas haloes.  We find the temperature of this region is consistent with the temperature of the gas at similar radii outside this bridge region.  Lastly, we examine the cluster dynamics using optical data.  We conclude A98N and A98S are likely bound to one another, with a 67\% probability, while A98S and A98SS are not bound at a high level of significance.
\end{abstract}


\section{Introduction} \label{intro}
The standard cold dark matter (CDM) model of structure formation predicts that clusters form in, and merge along, large scale structure filaments, with massive clusters located at the intersections of filaments. The fact that the three main subclusters in A98 lie along a line on the sky suggests the presence of such a filament.  These filaments trace the large-scale structure of the Universe, and are thought to contain most of the ``missing baryons" in the local Universe.  These ``missing baryons" are baryons that are known to exist at high-redshift (measured from primordial nucleosynthesis and the study of Ly$\alpha$ lines near $z=3$), but are not seen at low-redshift.  At low redshifts, only about 10\% of the expected baryons are found in galaxy clusters and groups.  The prevailing model argues that these missing baryons may be located in a low-density gas known as the warm-hot intergalactic medium (WHIM).  The WHIM has temperatures in the $kT\approx0.01-1$~keV range.  For a review, see \citet{bregman}.

Because of their low densities and low surface brightnesses, these filaments are difficult to observe directly.  There are few reports of direct detection; these have been of the denser, hotter part of the WHIM at or just beyond the virial radius of clusters~\citep{tittley, werner}.  Attempts have been made to observe filaments in X-ray absorption~\citep[e.g.][]{nicastro}, but these results are controversial~\citep{kaastra, rasmussen}.  A more targeted method for detecting the WHIM in X-ray absorption (using AGN as the background source) makes use of foreground structures with a known redshift that are likely to be associated with the WHIM.  This has led to stronger detections~\citep{buote, fang}.  

As clusters merge along these large scale filaments, they leave their imprint on the intracluster medium (ICM).  Recent X-ray observations have shown substructure in the ICM in the form of shocks, cold fronts, and gas sloshing spirals.  These all appear as surface brightness discontinuities in the X-ray images.  This substructure is the result of cluster-cluster (or group) interactions~\citep[and references therein]{markevitch}.

Shocks in clusters are consequences of large-scale cluster mergers, or sometimes of outbursts from the central active galactic nucleus (AGN). The temperature, density, and pressure all change across a shock front.  Typical cluster merger shocks have Mach numbers $M\lesssim3$~\citep{markevitch}.  The Bullet Cluster provides a textbook example of such a merger shock in the ICM~\citep{bullet}.  In cold fronts and sloshing spirals, only the density and temperature change across the surface brightness discontinuity, while the pressure does not, in contrast to shocks.  Also, in both merger and sloshing cold fronts the gas on the denser side of the front is cooler, whereas in shocks the denser gas is hotter~\citep{markevitch}.  While cold fronts are consequences of major and minor mergers, as well as gas sloshing, spirals are unambiguous markers of gas sloshing, which is triggered by an off-center minor merger that imparts angular momentum to the gas.  Such spirals are not one continuous spiral cold front, but rather individual cold fronts that combine into a spiral pattern.  As time goes by and the large scale structure grows, however, a coherent spiral emerges~\citep{am}.

Abell~98 is a ``triple'' cluster at a redshift of $z=0.1042$.  It is a Bautz-Morgan type II-III and a richness class 3~\citep{abell}.  There are three main components: A98N (J2000: $\alpha$:~00h46m25.0s, $\delta$:~20$^\circ$37$'$14$''$), with a redshift $z=0.1043$; A98S (J2000: $\alpha$:~00h46m29s, $\delta$:~20$^{\circ}$28$'$04$''$), with a redshift $z=0.1063$; and A98SS (J2000: $\alpha$:~00h46m36.0s, $\delta$:~20$^\circ$15$'$44$''$), with a redshift $z=0.1218$.  The projected distance between A98N and A98S is 1.1~Mpc, and the projected distance between A98S and A98SS is 1.4~Mpc.  Positions reported for A98N and A98SS are the positions of the peak of the \textit{Einstein} X-ray surface brightness reported by \citet{jones1999}, while the position reported for A98S is the position of the AGN~\citep{burns1994}.  The redshifts reported for A98N and A98S are from \citet{white}, while the redshift for A98SS was calculated using the velocities reported in \citet{pinkney}.  See \S\ref{dynamics} for more details.  A98 was studied extensively in the optical by \citet{fd} and \citet{dresslera, dresslerb}.  The original optical data were shown by \citet{henry} to have a bimodal distribution, which they designate as A98N and A98S. This bimodal distribution was also seen in early \textit{Einstein} X-ray observations of the ICM~\citep{bimodal, henry}.  A third component (A98SS) was seen in the X-ray by \citet{jones1999} using the \textit{Einstein} Observatory. 

A98S is host to the radio source 0043+2011 (4C 20.04).  Despite its straight morphology, it is considered a wide-angle tail (WAT) radio source since the jets disrupt and decollimate into tails~\citep{wat}.  It is possible that bending may be occurring along the line of sight, so that the jets appear straight in projection.  AGN jets can also interact with the ICM, evacuating cavities, which are seen in A2052~\citep{blanton} and Perseus~\citep{fabian}, for example.

This paper is organized as follows:  in \S\ref{obs:all}, we describe the X-ray observations and data reduction.  In \S\ref{spectra}, we discuss our analysis and fitting of spectra.  In \S\ref{mergersig}, we discuss the merger history and dynamical state of A98, and in \S\ref{feedback}, we discuss AGN feedback in A98S.  In \S\ref{filaments}, we investigate whether we are able to detect a filament.  Finally, in \S\ref{dynamics}, we discuss the cluster dynamics using optical data.  Throughout, we assume a cosmology with $H_0=70$~km~s$^{-1}$~Mpc$^{-1}$, $\Omega_{\Lambda}=0.7$, and $\Omega_M=0.3$.  At the redshift of A98 (z=0.1042), this gives a scale of $1''=1.913$~kpc and a luminosity distance $D_L=481.0$~Mpc.  All luminosity distances were calculated using the online Cosmology Calculator~\citep{cosmocalc}.  Unless otherwise stated, reported errors correspond to 90\% confidence intervals and analysis has been performed on the \textit{Chandra} data.

\section{Observations} \label{obs:all}

\subsection{\textit{Chandra}} \label{obs:chandra}
We used three \textit{Chandra} pointings:  two from 2009 September 17 (OBSID 11876 and OBDSID 11877, totaling 19.2~ks and 17.9~ks, respectively), and one from 2011 September 08 (OBSID 12185, totaling 18.7~ks).  Reported exposure times have been filtered for flares.  Observation information is summarized in Table~\ref{obs}.  Approximately 3.3~ks of time was filtered from the observations.  All three observations were taken in very faint (VFAINT) mode using the ACIS-I array.  The first two observations cover the subclusters A98N and A98S, while the third covers the subclusters A98S and A98SS.

The data were reprocessed from the Level 1 events files using CIAO version 4.5 and CALDB version 4.5.6.  The data were filtered to remove periods with strong flares using \texttt{lc\_clean}\footnote{\texttt{http://asc.harvard.edu/contrib/maxim/acisbg/}} and background data sets were created using the appropriate blank sky background fields reprojected to match the observations.  For more details on the reprocessing, see \citet{randall2011}.  The background fields were normalized to the observations in the 10--12~keV range.  Exposure maps for each dataset were created assuming an isothermal plasma emission model with $kT = 3$~keV, which is approximately the temperature of the subclusters listed in \citet{white}. 

\begin{figure}
\begin{center}
\plotone{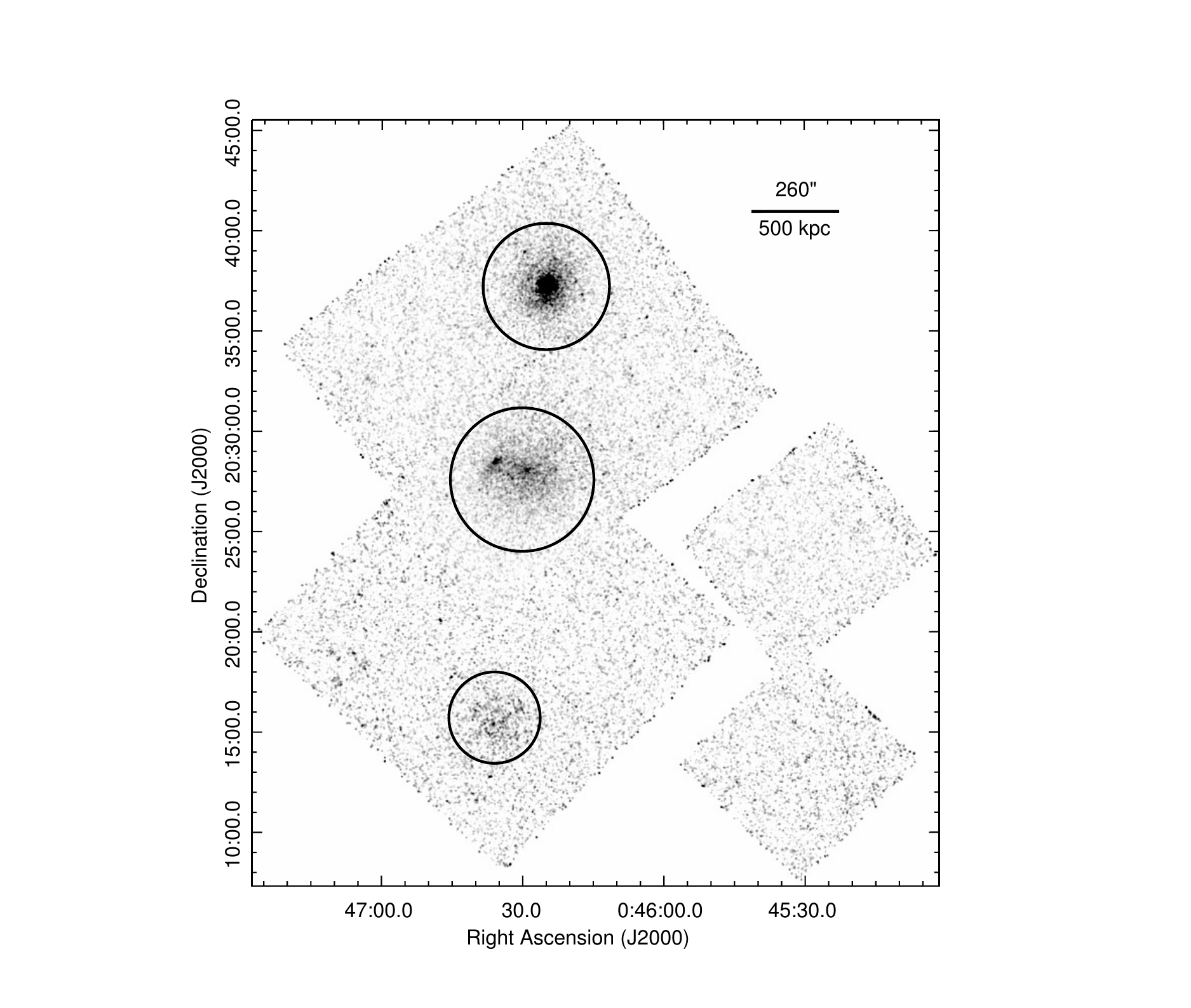}
\caption{Merged, background subtracted, and exposure map corrected \textit{Chandra} image of the extended, diffuse emission of A98 in the 0.3--8.0~keV energy band.  The image has been smoothed with a $5''$ Gaussian.  The subclusters are circled with the regions used for the spectral extraction.  From north to south they are: A98N, A98S, and A98SS.  A98N shows a regular distribution and a bright central peak, attributes indicative of a cool core cluster.  A98S and A98SS do not have regular morphologies or bright central cores, as is typical of non-cool core clusters.}
\label{a98}
\end{center}
\end{figure}
Figure~1 shows a merged, background- and exposure-corrected image of the extended, diffuse emission in the energy range 0.3--8.0~keV.  Point sources were detected using the CIAO wavelet detection tool \texttt{wavdetect}.  Wavelet scales of 1, 2, 4, 8, and 16 pixels were used, and the significance threshold was set to $10^{-6}$.  The point sources were detected using an unbinned, unsmoothed image (pixel scale of $0\farcs492$) for the individual datasets and then merged.  This was done a second time on an image with the previous point sources removed in order to detect fainter sources.  We visually examined the detected sources to remove duplicates, as well as remove detections that were likely to be structure in the diffuse gas.  The resulting point sources were removed, and the excluded regions were filled using the surface brightness of the surrounding regions and the CIAO tool \texttt{dmfilth}.  

In Figure~\ref{a98}, the three subclusters are clearly visible.  A98N, the northernmost subcluster, has the morphology of a cool core cluster, and, as we will discuss in \S\ref{mergersig:a98n}, there is evidence of either a gas sloshing spiral or a shock in the ICM.  A98S, the central subcluster, has a disturbed morphology, with two bright peaks.  One of these peaks is the central point source corresponding to the AGN that is powering the WAT.  There is also a diffuse peak to the east of the AGN.  A98SS, the southernmost subcluster, also has a disturbed morphology, with no bright peaks.  We discuss the dynamical state of, and merger signatures in, each of the subclusters in \S\ref{mergersig}.
  
\subsection{\textit{XMM-Newton}} \label{obs:xmm}

A98 was observed with \xmm with two observations: July 2010 (OBSID  0652460101; with filtered exposure times of 11.7~ks for MOS1, 16.9~ks for MOS2, and 3.2~ks for PN) centered on A98S and A98SS and December 2010 (OBSID 0652460201; with filtered exposure times of 27.9~ks for MOS1, 29.8~ks for MOS2, and 21.4~ks for PN) centered on A98N and A98S.  The observations are summarized in Table~\ref{obs}. Both datasets were processed using the \xmm Extended Source Analysis Software (\esas, \citealp{snowden}) in combination with the Science Analysis System (SAS) version 13.0.3, and the most recent calibration files as of September 2013. The details of the \xmm analysis are described in \citet{bulbul2012}.  We summarize the main analysis methods here.
The calibrated clean event files were produced after filtering for the high intensity soft proton flares. 

\begin{figure}
\begin{center}
\plotone{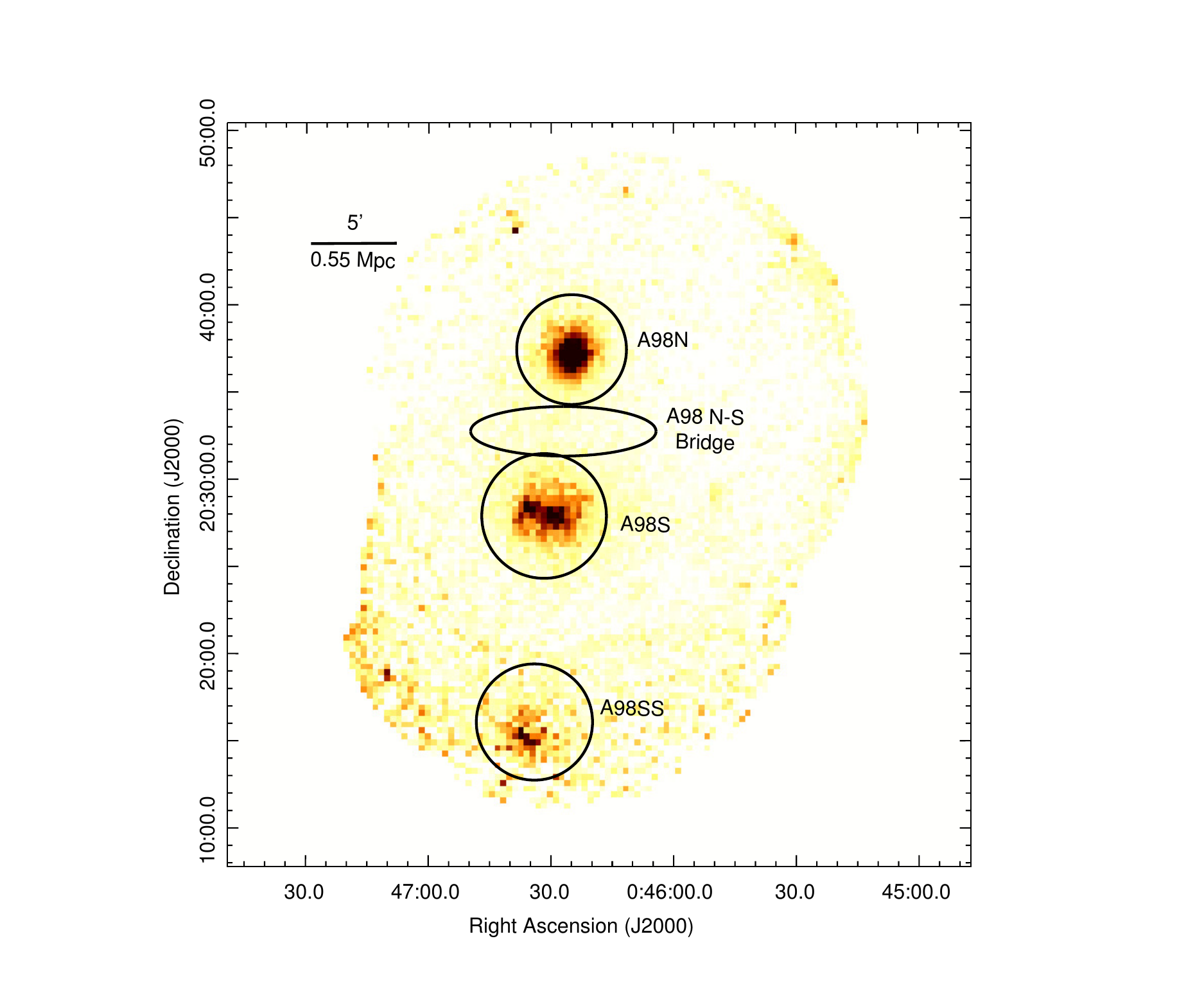}
\caption{ Exposure and background corrected  \xmm EPIC image of the triple cluster system A98 in the energy band 0.4--10.0~keV. The image has been binned.
 The circles and ellipse show the regions from which the \xmm spectra were extracted. }
\label{fig:a98-xmm}
\end{center}
\end{figure}

Images were created in the $0.4-7.0$~keV band for the MOS and PN detectors. 
Identical regions corresponding to the point sources detected in \textit{Chandra} were removed from the \xmm images, and were visually checked to compare with the point sources detected with \xmm. The images were examined 
carefully to check for individual CCDs operating in an anomalous state.  In both observations, MOS1 CCD4 was operating in an anomalous state and thus eliminated from further analysis.  Additionally, MOS1 CCD6 was eliminated from analysis due to a micrometeorite hit.
Figure \ref{fig:a98-xmm} shows the full combined mosaic \xmm MOS and PN image of A98.
\begin{deluxetable*}{lcccccc}
\tablecaption{X-Ray Observations}
\tablewidth{0pt}
\tablehead{\colhead{OBSID} & \colhead{Observatory} & \colhead{Start Date} & \colhead{Exp}\tablenotemark{a} & \colhead{PI} & \colhead{Type}\\
\colhead{} & \colhead{} & \colhead{} & \colhead{(ks)} & \colhead{} & \colhead{}}
\startdata
11876 & \textit{Chandra} & 2009-09-17 & 19.2 & S. Murray & GTO\\
11877 & $\textit{Chandra}$ & 2009-09-17 & 17.9 & S. Murray & GTO\\
12185 & \textit{Chandra} & 2010-09-08 & 18.7 & S. Murray & GTO\\
0652460101 & \xmm & 2010-07-26 & 11.7/16.9/3.2 & C. Jones & GO\\
0652460201 & \xmm & 2010-12-26 & 27.9/29.8/21.4 & C. Jones & GO \\
\enddata
\tablenotetext{a}{Exposure times have been filtered.  For \xmm exposures, they are listed MOS1/MOS2/PN.}
\label{obs}
\end{deluxetable*}

\section{Spectral Analysis} \label{spectra}
\subsection{Total Spectrum}
We examined the spectra of each subcluster to determine the global properties of the ICM, using the same regions for both \textit{Chandra} and \xmm. These regions were chosen by eye and incorporate the obvious extended emission.  In all cases the absorption was fixed at the weighted average Galactic value of $N_H=3.06\times10^{20}$~cm$^{-2}$ calculated in the LAB survey~\citep{nh}. Varying the absorption did not significantly improve the spectral fit in any case.  The XSPEC default solar table from \citet{solar} was used.   The results from the spectral fitting are summarized in Table~\ref{fits}.
 
\subsubsection{\textit{Chandra}} \label{spec:chandra}
The spectral analysis was performed with point sources excluded, and, unless otherwise noted, the spectra were binned with a minimum of 40~counts per bin.  The regions used for extraction are shown in Figure~\ref{a98}.  The spectra from the different observations were fitted simultaneously using the XSPEC software package, version 12.7.0.  For details on XSPEC, see \citet{xspec}.  The fitting was done in the energy range 0.6--7.0~keV, and the blank-sky background files were scaled so that the 10--12~keV count rate matched that of the observations to account for the quiescent particle background.  For each subcluster, we used an absorbed single-temperature collisional equilibrium plasma (APEC) model~\citep{apec}, which utilizes AtomDB v2.0.1~\citep{foster}.  
\begin{deluxetable*}{lccccc}
%\tabletypesize{\footnotesize}
\tablecaption{XSPEC Fits to the Three Subclusters in A98}
\tablewidth{0pt}
\tablehead{\colhead{Subcluster} & \colhead{Observatory} & \colhead{$kT$} & \colhead{Abundance} & \colhead{$\chi^2/dof$} & \colhead{No. of Counts}\\
\colhead{} & \colhead{} & \colhead{(keV)} & \colhead{($Z_{\sun}$)} & \colhead{} & \colhead{}}
\startdata
A98N & \textit{Chandra} & $3.07^{+0.21}_{-0.21}$ & $0.38^{+0.12}_{-0.10}$ & $190/207=0.92$ & 7600\\
A98N & $\xmm$ & $3.09^{+0.19}_{-0.20}$ & $0.56^{+0.08}_{-0.09}$ & $1866/1912=0.98$ & 8300/9200/13500\tablenotemark{a}\\
A98S & \textit{Chandra} & $2.77^{+0.22}_{-0.21}$ & $0.22^{+0.09}_{-0.08}$ & $306/294=1.04$ & 8900\\
A98S & \xmm & $3.03^{+0.37}_{-0.31}$ & $0.27^{+0.06}_{-0.07}$  & $598/536=1.12$ & 2600/4750/6300\tablenotemark{b}\\
A98SS & \textit{Chandra} & $2.85^{+0.58}_{-0.55}$ & $0.45^{+0.43}_{-0.27}$ & $44/41=1.08$ & 1200\\
A98SS & \xmm & $3.07^{+0.87}_{-0.06}$ & $0.34^{+0.38}_{-0.18}$ & $1254/1448=0.87$ & 2000/2500\tablenotemark{c}\\
\enddata
\tablenotetext{a}{Counts are listed MOS1/MOS2/PN.  All data comes from OBSID 0652460201.}
\tablenotetext{b}{Counts are listed MOS1/MOS2/MOS2.  The first two are for OBSID 0652460101, while the third is for OBSID 0652460201.}
\tablenotetext{c}{Counts are listed MOS1/MOS2.  Both come from OBSID 0652460101.}
\label{fits}
\end{deluxetable*}

We first fitted each of the subclusters with a single APEC model with the absorption fixed to $3.06\times10^{20}$~cm$^{-2}$.  For the northern subcluster, this yields a temperature of $kT = 3.07\pm0.21$~keV and an abundance of $Z = 0.38^{+0.12}_{-0.10}$~$Z_{\sun}$.  For A98S, we found a temperature of $kT = 2.77^{+0.22}_{-0.21}$~keV and an abundance of $Z = 0.22^{+0.09}_{-0.08}$~$Z_{\sun}$.  In A98SS, the temperature is $kT = 2.85^{+0.58}_{-0.55}$ and the abundance is $Z = 0.45^{+0.43}_{-0.27}$.  For all three subclusters, the temperatures and abundances are consistent within the error bars.

We also used a single APEC model with $N_H$ allowed to vary.  With each subcluster we performed an F-test to check whether leaving the $N_H$ free to vary improved the results.  For A98N and A98SS there was no significant improvement in the fit when absorption was free to vary.  In A98S, the F-test yielded a probability value of  $\sim3\%$; however the fit to $kT$ was insensitive to $N_H$ and the $N_H$ value derived was consistent with zero.  Allowing $N_H$ to vary did not give better results.  Therefore, throughout the rest of the paper, we fix $N_H$ at the Galactic value.
\\
\subsubsection{\textit{XMM-Newton}} \label{spec:xmm}

Spectra were extracted from the \xmm observations using the \sas tools \textit{mos-spectra} and \textit{pn-spectra}.
Redistribution matrix files (RMFs) and ancillary response files (ARFs) were created with the \sas tools \textit{rmfgen} and \textit{arfgen}, respectively. 

Each extracted spectrum includes contributions from the quiescent particle background (QPB); soft X-ray background emission including solar wind charge exchange (SWCX), Galactic halo (GH), local hot bubble (LHB), and extragalactic emission; and residual contamination from soft protons.  QPB spectra were extracted from the filter wheel closed (FWC) data.  The details of the background analysis and subtraction can be found in \citet{bulbulxmm}.

Energies outside the range 0.3--10.0~keV were ignored and all EPIC-MOS and PN spectra were 
fitted simultaneously.  We included the high-energy (7.0--10~keV) end of the spectrum  to fit the power-law slope and normalization, which are determined by the high end of the spectrum.  This power-law is used to eliminate the soft proton residuals.  The diffuse X-ray emission was fitted with an APEC model using $\chi^2$-statistics. The temperature and abundance measurements for each cluster are summarized in Table \ref{fits}.  The temperatures for all three subclusters measured using \xmm are consistent with the temperatures measured using \textit{Chandra}, as are the abundance measurements for the three subclusters.  Results from \xmm indicate a higher abundance in A98N as compared with A98S (the constraints on the abundance for A98SS are poor).  Although the measured abundances are within the range reported for clusters and groups~\citep[e.g.,][]{maughan}, it is possible that the apparent high abundance in A98N is due to the ``inverse Fe bias''~\citep{gastaldello}.  This effect artificially elevates the abundance measurements when modeling multi-temperature emission from clusters in the temperature range $2-4$~keV with a single temperature model.  There were insufficient counts to constrain multi-temperature emission models.  

\subsection{X-Ray Temperature Maps} \label{tmaps}
A spectral map was created from the \textit{Chandra} data to examine the temperature distribution in the ICM, following the technique outlined in \citet{randall2008,randall2009}.  Point sources were excluded for the analysis.  For each pixel in the temperature map, spectra were extracted from a surrounding circular region that contained roughly 1200~net counts.  Because of this, each region of spectral extraction has a different size, the smallest of which has a radius of $\sim18''$.  The spectra were extracted in the 0.6--7.0~keV energy range, with the column density fixed to $N_H = 3.06\times10^{20}$~cm$^{-2}$.  The abundance was allowed to vary.

\begin{figure}
\begin{center}
\plotone{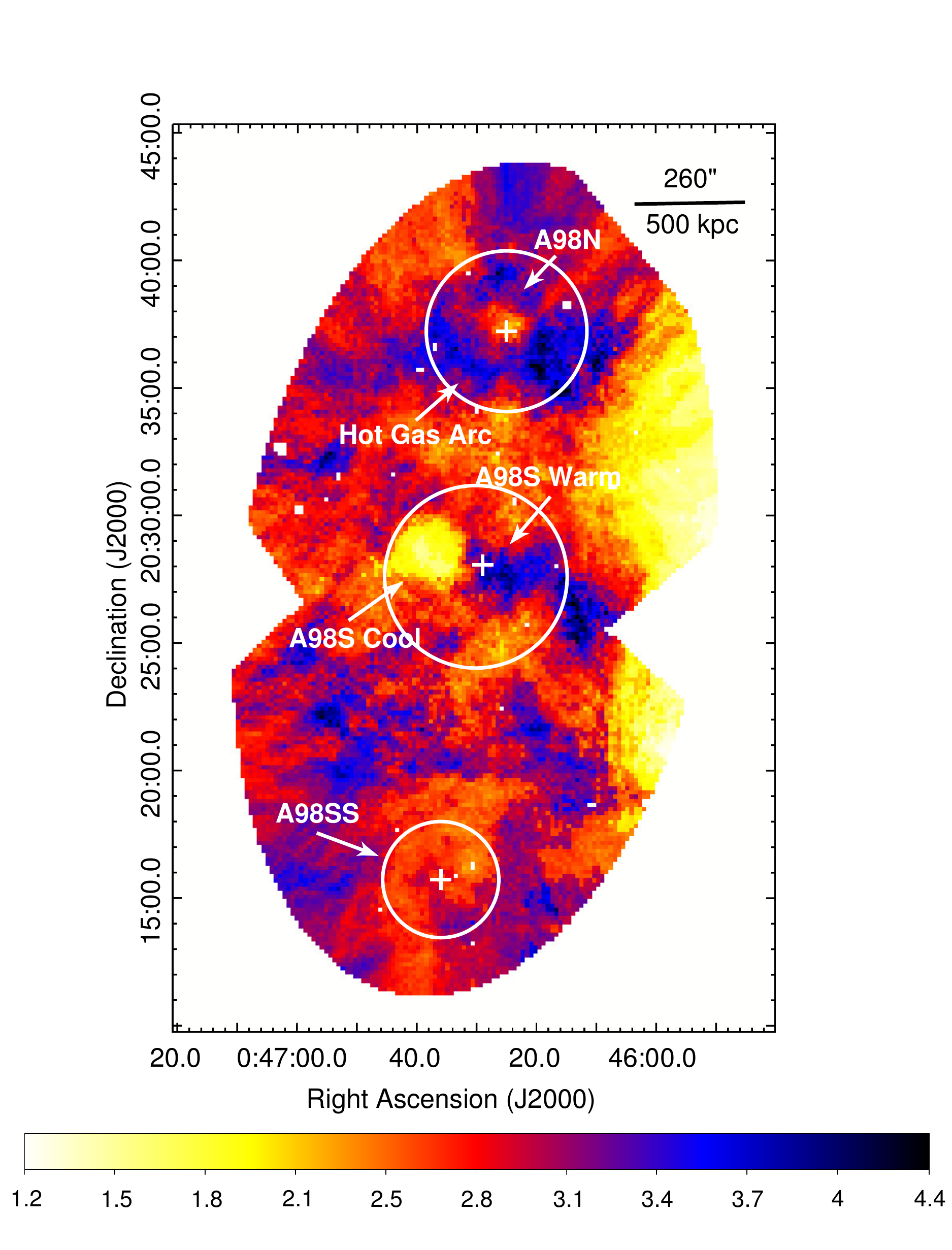}
%\includegraphics[scale=0.73]{tmap_new.pdf}
%\vspace{-5ex}
\caption{Temperature map of A98 with the scale bar in units of keV.  The subclusters' centers (as listed in \S\ref{intro}) are marked by crosses and circled with the same regions as in Figure~\ref{a98}.  A98N has a cool core with a ring of warmer gas surrounding it, particularly to the south, possibly due to shock heated gas. A98S has an asymmetric temperature structure, with a local temperature minimum to the east and a local maximum to the west.  A98SS is also marked by the circle (using the same region shown in Figure~\ref{a98}).}
\label{tmap}
\end{center}
\end{figure}

Figure~\ref{tmap} shows the temperature map of A98.  A98N has a cool core with a ring of warmer gas surrounding it, which is indicative of a cool core cluster.  The cool core gas temperature is $kT\sim2.3$~keV, while the surrounding gas temperature is $kT\sim3.6$~keV.  The errors on the temperature map values are approximately 15-20\% in the brighter regions, and go as high as $\sim45$\% in the fainter regions away from the subcluster cores.  The warmer gas appears to be asymmetric, which could indicate the presence of cooler ``sloshed" gas to the north or a merger shock to the south.  A98S has an asymmetric temperature structure, with an area of warmer gas to the west that is coincident with the WAT AGN.  The cooler gas temperature is $kT\sim2.0$~keV, while the warmer gas temperature is $kT\sim3.8$~keV.  The chip containing A98SS has a low number of counts.  Since each region is allowed to grow until it contains 1200 counts, the regions are larger and thus the map is highly smoothed.  The low number of counts also means that it is difficult to see the subcluster's structure in the temperature map.  There is, however, a large region where the temperature is $kT\sim2.7$~keV, which is consistent with the temperature of $kT = 2.65$~keV measured from extracting the global spectrum for A98SS using the extraction region shown in Figure~\ref{a98}.  There does not appear to be a cool core.  The features in A98N and A98S described above will be examined in more detail in \S\ref{mergersig}.

\section{Subcluster Dynamical States and Merger Signatures} \label{mergersig}
\subsection{A98N} \label{mergersig:a98n}

Figure~\ref{tmap} shows that A98N has the characteristics of a cool core cluster, but that the surrounding warmer gas is distributed asymmetrically, consistent with either cooler core gas displaced to the north by gas sloshing or the presence of a merger shock to the south.  Here, we examine whether the subcluster shows evidence for either of these scenarios.

\begin{figure}
\begin{center}
\plotone{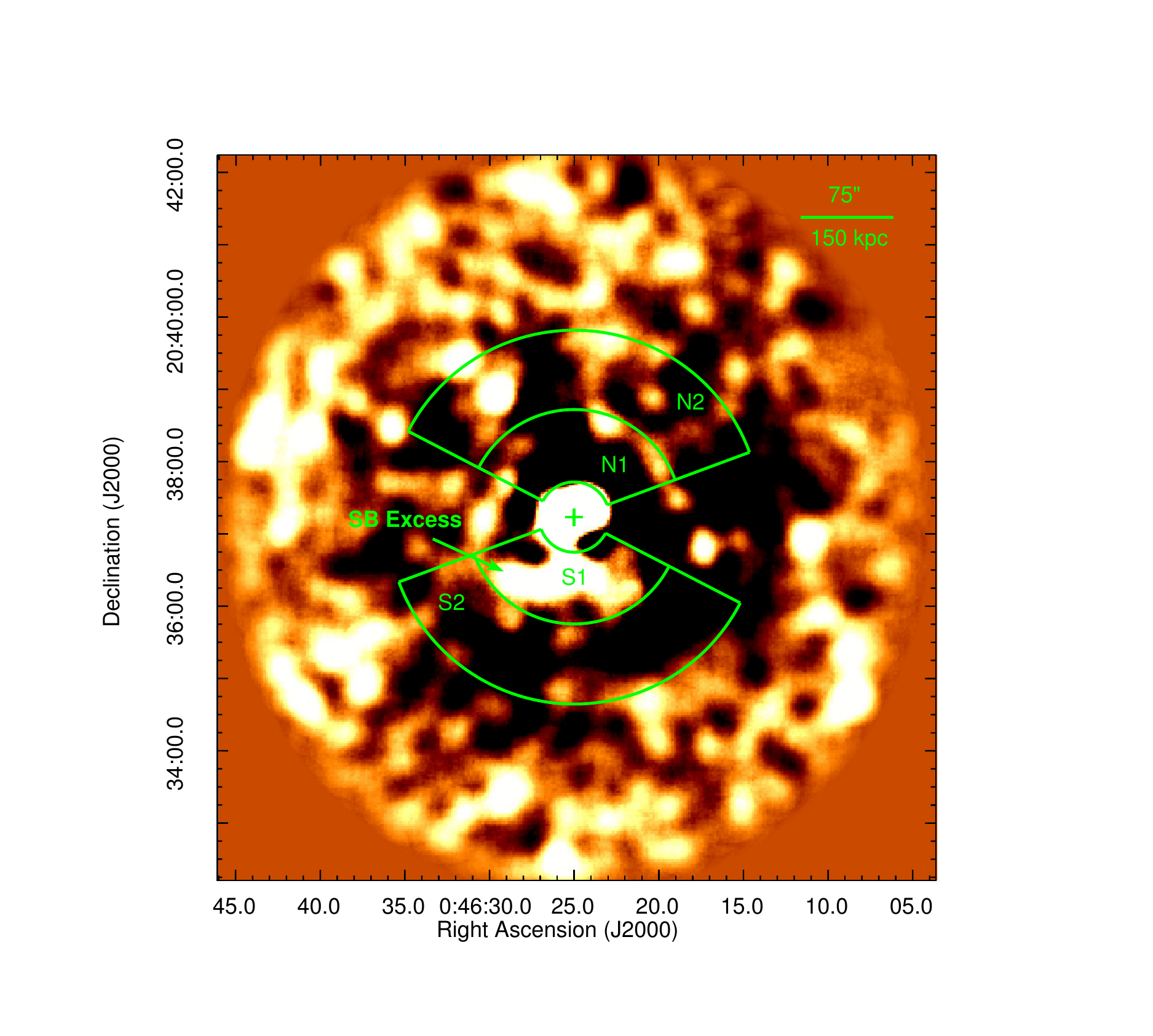}
\caption{Beta model subtracted \textit{Chandra} image of A98N.  The cross marks the center of the subcluster.  There is an asymmetrical surface brightness excess.  The surface brightness excess (S1) is spatially coincident with the hotter gas surrounding the cool core of A98N, which indicates that it could be a shock.  The image has been smoothed with a 20" Gaussian.  The labeled regions are described in detail in the text.}
\label{a98n_beta}
\end{center}
\end{figure}

We first made a beta-model subtracted \textit{Chandra} image to search for faint substructure in the diffuse emission, e.g., a gas sloshing spiral or a merger shock, from A98N.  A 2D elliptical beta-model was used to fit the surface brightness of the point source-free 0.3--10~keV image, which was binned by a factor of four.  The model was fitted in \textit{Sherpa} using Cash statistics, and corrections were made for exposure using the merged exposure map.  A constant background rate of $5.1\times10^{-8}$~counts~s$^{-1}$~arcsec$^{-1}$ was applied.  The model was subtracted from the original image, leaving the residual image shown in Figure~\ref{a98n_beta}.  The fit was dominated by emission from the bright core, and the model was not able to describe the fainter extended emission well.  For the purpose of highlighting asymmetries in the central region, however, the model is useful.

In Figure~\ref{a98n_beta}, there is a clear surface brightness excess directly south of the subcluster core, which appears to curve clockwise to the north of the core.  The curvature of this surface brightness feature is indicative of a gas sloshing spiral.  Such spirals were originally simulated by \citet{am}.  They have been observed in clusters such as Perseus~\citep{perseus, fabian}, Virgo~\citep{virgo}, A496~\citep{a496}, A2029~\citep{a2029, rpm}, as well as many others~\citep{lagana}.  For a review, see \citet{markevitch}.  This surface brightness excess is coincident with the warmer gas that surrounds the cluster core seen in Figure~\ref{tmap}, however, which is suggestive of a shock. 

To determine whether or not the surface brightness excess is a spiral or a shock (or neither), we used a single APEC model in XSPEC to calculate the temperature of two wedges in either direction--to the north of the core and to the south of the core.  Figure~\ref{a98n_beta} shows the regions used.  One of the regions to the south of the core (S1) contained the surface brightness excess.  The S1 wedge has a temperature of $kT = 4.22^{+1.48}_{-0.95}$~keV.  The temperature of S2 is $kT = 3.36^{+0.50}_{-0.40}$~keV.  To the north, the N1 wedge has a temperature of $kT = 3.28^{+0.92}_{-0.63}$~keV.  N1 is at the same radius as S1.  The N2 wedge has a temperature of $kT = 3.36^{+0.65}_{-0.48}$~keV, which is consistent with the temperature of S2.  The S1 wedge is warmer than both the S2 and N1 wedges at the $1.4\sigma$ level, and warmer than N2 at the $1.3\sigma$ level.  Since the region with the surface brightness excess has a higher temperature (albeit at a low significance level) than the three other regions, which all have comparable temperatures of $\sim3.3$~keV, and the location of the surface brightness excess is located where one would expect for a leading bow shock due to a merger with A98S, these results are more consistent with a shock than with a gas sloshing spiral.

Using the temperature in the S1 wedge and the temperature in the S2 wedge we are able to calculate a Mach number for a possible shock occurring in this region.  We find that $M=1.3^{+0.6}_{-0.3}$.  While these estimates are very rough, they are reasonable for a major merger, where Mach values are typically between 1 and 3~\citep{markevitch}.  If this temperature difference is due to a shock, then the Mach number given here is a lower limit, since projection effects will decrease the apparent temperature increase at the shock.

We also tried to fit both surface brightness and temperature profiles across the region.  We were unable to see an edge in the surface brightness profile.  In \S\ref{dynamical}, we calculate that the projection angle of the A98N-A98S system with respect to the sky is either $67.1^\circ$ or $25.2^\circ$.  Since the merger is not in the plane of the sky, the leading shock front edge will be blurred, making it difficult to detect an edge while still allowing a potential detection of shock heated gas.  We were unable to create a temperature profile of the region, as there were too few counts.

The potential shock-heated material is $\sim150$~kpc, in projection, to the cluster core of A98N, significantly closer than A98S, which is at a projected distance of 1.1~Mpc.  As mentioned previously, our dynamical analysis suggests that the merger is not in the plane of the sky, which would increase the actual separation between the putative shock and the core of A98N (although this also means that A98S is farther away from the shock).  We are unaware of any unambiguous detections of such early stage merger shocks in the literature that can be compared with our result.  Numerical simulations would be useful to confirm shock formation close to the core in an early stage merger; however they are beyond the scope of this work.

\subsection{A98S} \label{mergersig:a98s}

Figure~\ref{tmap} shows that there is a dual temperature structure in A98S:  there is a cool region to the east and a warmer region to the west.  The warmer region is coincident with the AGN in A98S, which suggests that it could be heating up the gas via AGN feedback.   Feedback from the AGN occurs when the jets of the AGN plow into the ICM.  The jets evacuate cavities in the ICM, creating bubbles, which have been seen in X-ray images of a large number of groups and clusters (e.g. A2052,~\citealp{blanton}; Perseus,~\citealp{fabian}; Hydra A,~\citealp{hydra}; and many others,~\citealp{birzan}).  In doing this, energy is transferred into the cooling gas, raising its temperature.  As the gas heats up, the accretion rate onto the central supermassive black hole (SMBH) drops, and the AGN jet power decreases.  As a result, the ICM heating rate drops, allowing the gas to cool and once again accrete onto the SMBH, and the cycle starts anew.  For a more detailed review, see \citet{feedback}.  We examined the two different regions individually using a single-temperature absorbed APEC model in XSPEC.  These regions were chosen to match the temperature structure in Figure~\ref{tmap}, and are shown in Figure~\ref{a98s_beta}.  Including the central point source that corresponds to the AGN in the analysis yields a temperature of $kT = 3.37^{+0.57}_{-0.39}$~keV.  If we exclude the central point source, the temperature of that same region is $kT = 3.41^{+0.65}_{-0.47}$~keV.  Thus, the central AGN does not significantly affect the results of the spectral fits in this region.  The eastern region has a temperature of $kT = 2.06^{+0.28}_{-0.13}$~keV, which is cooler than the western region at the $3\sigma$ level.

We also examined the central point source individually, which is expected to have a non-thermal power-law spectrum.  There were approximately 70~counts in the region of the point source in the 0.6--7~keV energy band, and thus we could not fit its spectrum to distinguish thermal from non-thermal emission.  Assuming that this central point source is an AGN (which is likely, given that it is at the center of the WAT), we can make an estimate of the luminosity.  We assume a photon index of $\Gamma=1.8$, which is the photon index seen in 90\% of radio-loud AGN~\citep{gamma}.  Using WebPIMMS\footnote{\texttt{http://heasarc.nasa.gov/Tools/w3pimms.html}}, we find that the X-ray luminosity of the point source is $9.0\times10^{41}$~erg~s$^{-1}$,  which is consistent with the luminosity of a low-luminosity AGN seen in weak-line radio galaxies~\citep{gamma}.

\begin{figure}
\begin{center}
\plotone{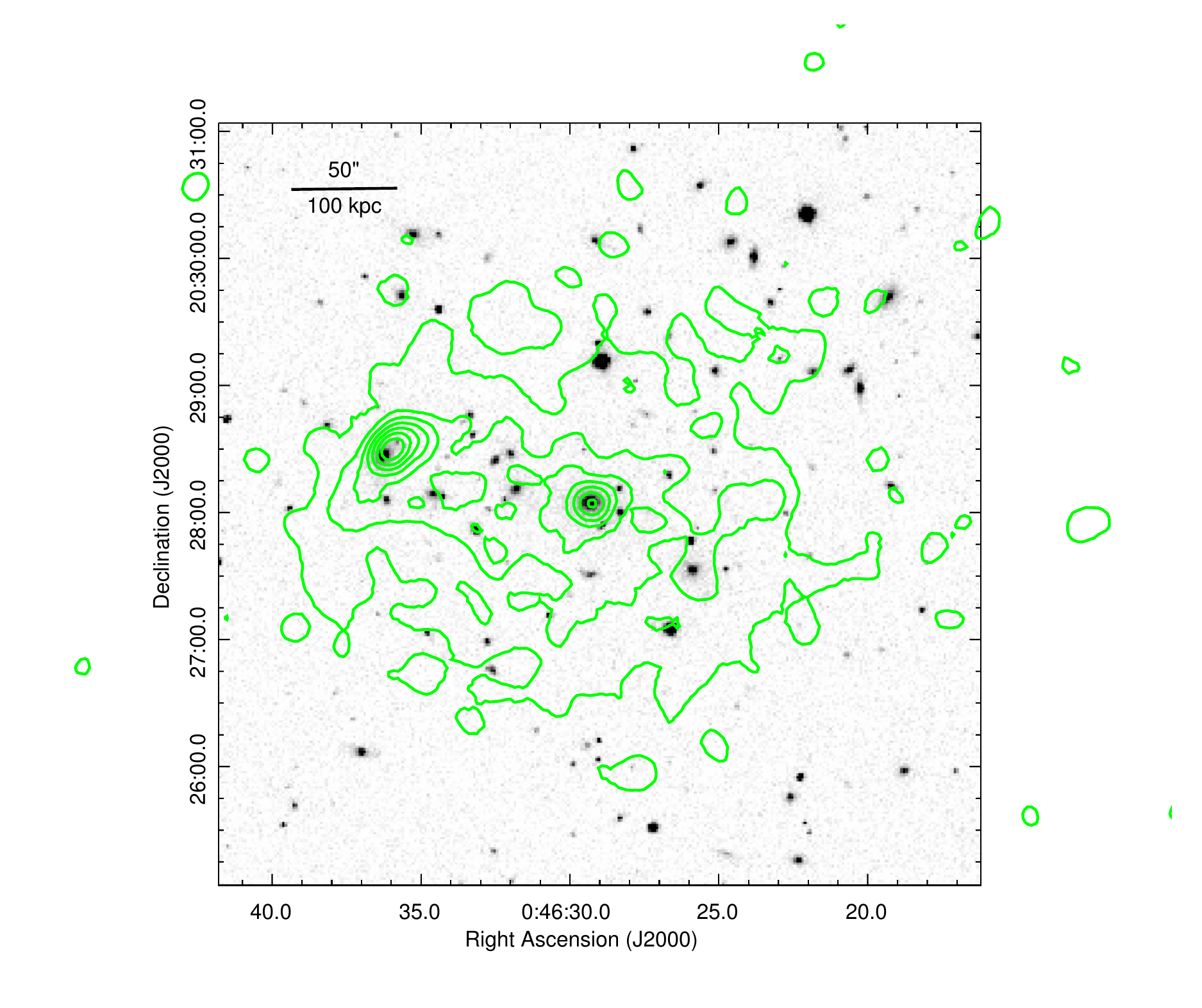}
\caption{X-ray contours from the \textit{Chandra} data overlaid on an SDSS-r image.  The two X-ray peaks are roughly spatially coincident with optical galaxies.}
\label{gals}
\end{center}
\end{figure}

The two temperature structure observed could be due to the AGN heating up the gas in the warmer section of the cluster; however, it is more likely that the two distinct temperatures seen are the result of a merger between two distinct subclusters. There are two peaks in the surface brightness, which can be seen in Figure~\ref{a98}.  One of the peaks is spatially coincident with the AGN, which has an optical galaxy counterpart with an SDSS-r magnitude of 15.14.  This is an SDSS Model magnitude, as are all other reported SDSS magnitudes.  This peak coincides with the warmer gas, as discussed earlier in this section.  The second surface brightness peak is roughly coincident with an optical galaxy that has an SDSS-r magnitude of 15.74.  The spatial coincidence between the X-ray surface brightness peaks and the optical galaxies can be seen in Figure~\ref{gals}.  
The next brightest cluster galaxy has an SDSS-r magnitude of 16.50.  The difference of nearly 0.75 mag between the second and third brightest cluster member and the fact that brightest cluster galaxies (BCGs) have a narrow spread of absolute magnitudes~\citep{sandage} suggest that this second surface brightness peak is a second BCG from a cluster merger.  The photometric redshifts from SDSS give a velocity difference of $\Delta v=300$~km~s$^{-1}$ for the two BCGs, which is reasonable for a recent merger; however the errors are very large: on the order of $\pm3000$~km~s$^{-1}$.  This merger could be causing dynamical heating, leading to the observed temperature structure.  It is also possible that the western subcluster was more massive, which would make it slightly warmer, or that the eastern subcluster could have been a cool core cluster, and the core has mostly survived the merger.  

\subsection{A98SS} \label{mergersig:a98ss}

While there are not enough counts in the region of A98SS to reveal detailed substructure or to map the temperature distribution, Figure~\ref{a98} does show that the gas distribution is quite diffuse, especially when compared to A98N.  All three subclusters have approximately the same mass (see \S\ref{dynamics}) and are all approximately at the same redshift.  If A98SS were a cool core cluster, we would likely be able to see the bright core, despite the exposure time being half as long. 

Figure~\ref{a98} shows that A98SS has a disturbed morphology.  Our dynamical analysis (see \S\ref{dynamics}) shows, however,  that A98SS is not bound to the other two clusters and thus its morphology cannot be caused by a merger with A98S.  Thus, it is possible that it has recently experienced a roughly 1:1 merger and is in the later stages of relaxing.  The cluster member identified by \citet{pinkney} as the BCG has an SDSS-r magnitude of 15.62.  The next brightest cluster member has an SDSS-r magnitude of 15.92.  It is possible that this is a second BCG, as after these cluster members the third brightest member has an SDSS-r magnitude of 16.60.  As in A98S, the two potential BCGs are very close in magnitude, and here, the difference between the second and third brightest cluster member is 0.68~mag.  Combined with our inability to see a cool core and the diffuse appearance of A98SS, this suggests that it has recently undergone a major merger.

\section{AGN Feedback in A98S} \label{feedback}

As described in \S\ref{mergersig:a98s}, the jets from the AGN can evacuate cavities in the ICM.  If the viewing angle is favorable, these cavities can be seen as deficits in the X-ray emission coincident with the radio lobes of the AGN.  Because there is a WAT AGN in A98S, we searched for cavities in the X-ray.
\begin{figure}
\begin{center}
\plotone{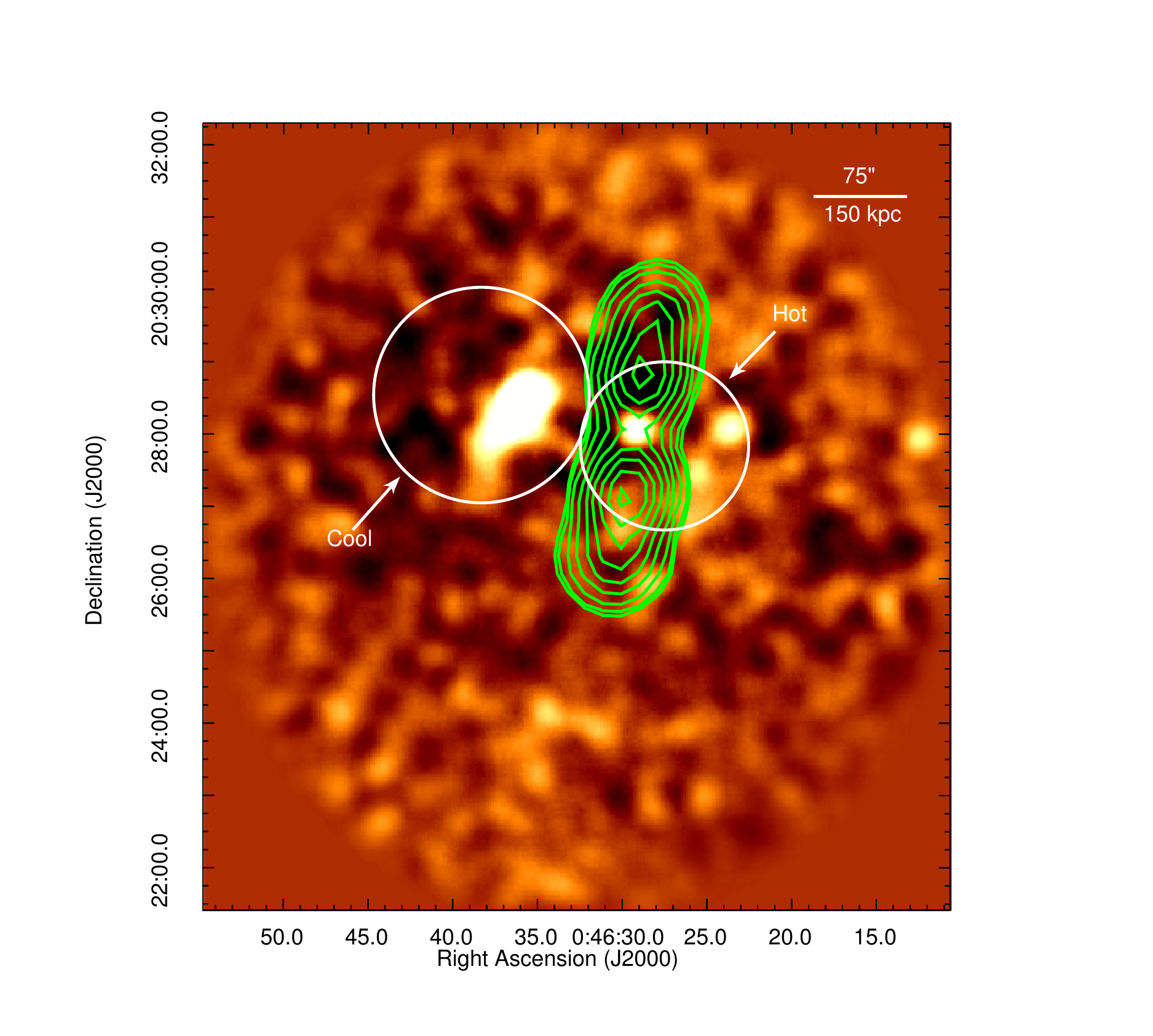}
\caption{Beta model subtracted \textit{Chandra} image of A98S with 1.4 GHz NVSS radio contours of the WAT AGN overlaid in green.  There is a surface brightness deficit under the northern lobe of the AGN, indicating that a cavity has been evacuated.  There is also a surface brightness excess to the east of the AGN, which appears to be the remnant core of another subcluster from an ongoing merger.  The image has been smoothed with a 20" Gaussian.  The regions labeled ``Hot" and ``Cool" were used for the spectral fits presented in \S\ref{mergersig:a98s}.}
\label{a98s_beta}
\end{center}
\end{figure}
Figure~\ref{a98s_beta} shows a beta-model subtracted residual image of A98S, constructed in the same manner as the beta-subtracted image for A98N (Figure~\ref{a98n_beta}).  As before, a constant background rate of $5.1\times10^{-8}$~counts~s$^{-1}$~arcsecond$^{-1}$ was applied.  The 1.4 GHz radio contours of the WAT are overlaid in green, and are from the NRAO VLA Sky Survey~\citep[NVSS;][]{NVSS}. 

There is an apparent surface brightness deficit under the northern radio lobe, indicating that the AGN is evacuating a cavity in the ICM.  There is also a region of excess surface brightness to the east, coincident with the eastern cold spot seen in the temperature map, which is most likely the result of a merging subcluster, as discussed in \S\ref{mergersig:a98s}. 

\begin{figure}
\begin{center}
\plotone{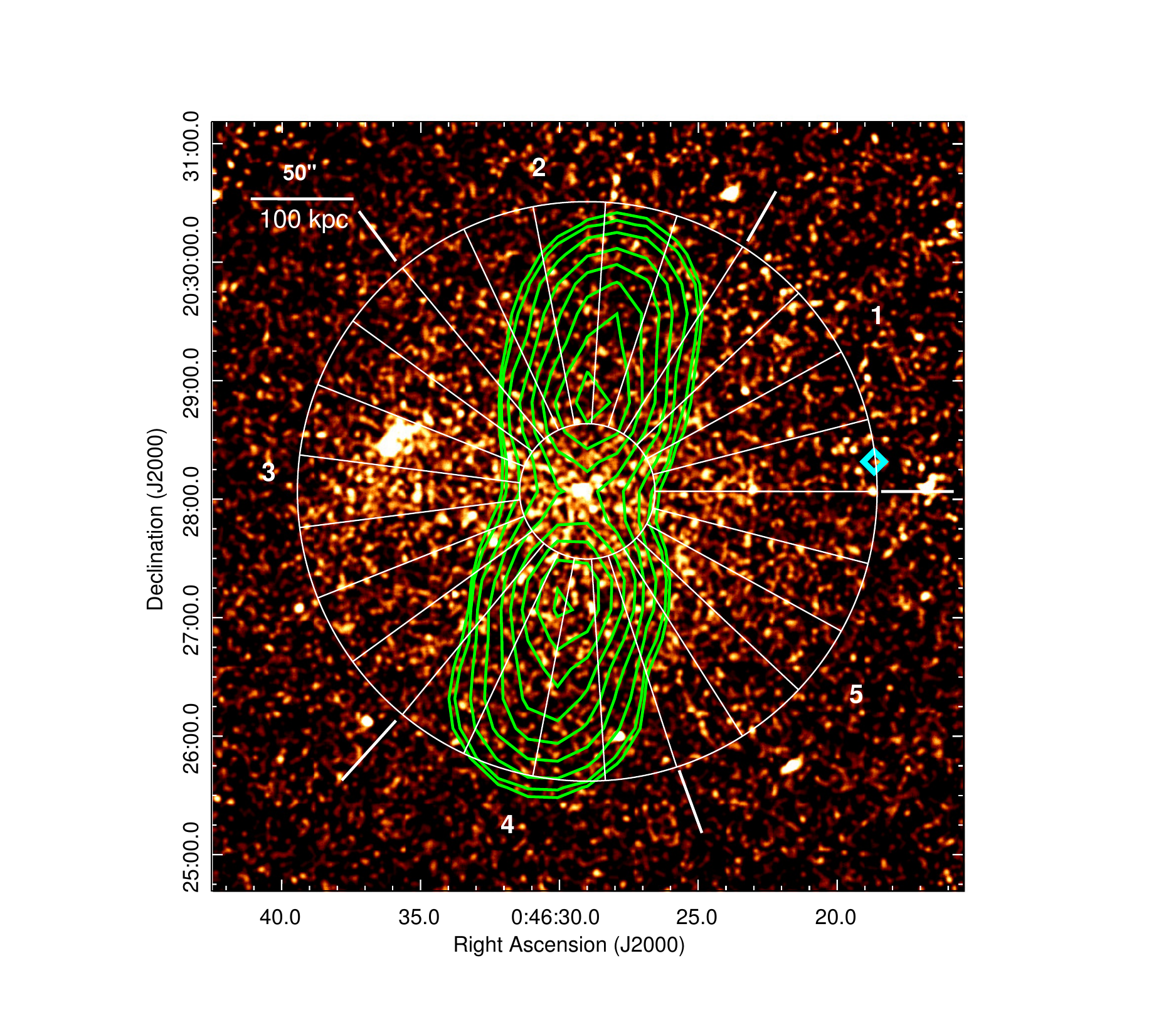}
\caption{A lightly smoothed ($2\farcs5$ Gaussian) \textit{Chandra} image of A98S with radio contours overlaid in green and the wedges used to make the azimuthal surface brightness profile overlaid in white.  The sections used for binning are also labeled.  Angles are measured counter-clockwise from the wedge marked with the aqua diamond.}
\label{regions_wat}
\end{center}
\end{figure}

To determine the significance of the surface brightness deficit, we made an azimuthal surface brightness profile in the 0.3--10~keV energy band using the regions shown in Figure~\ref{regions_wat}.  Point sources were excluded.  Angles are measured counter-clockwise from the wedge marked with the diamond.

\begin{figure}
\begin{center}
\plotone{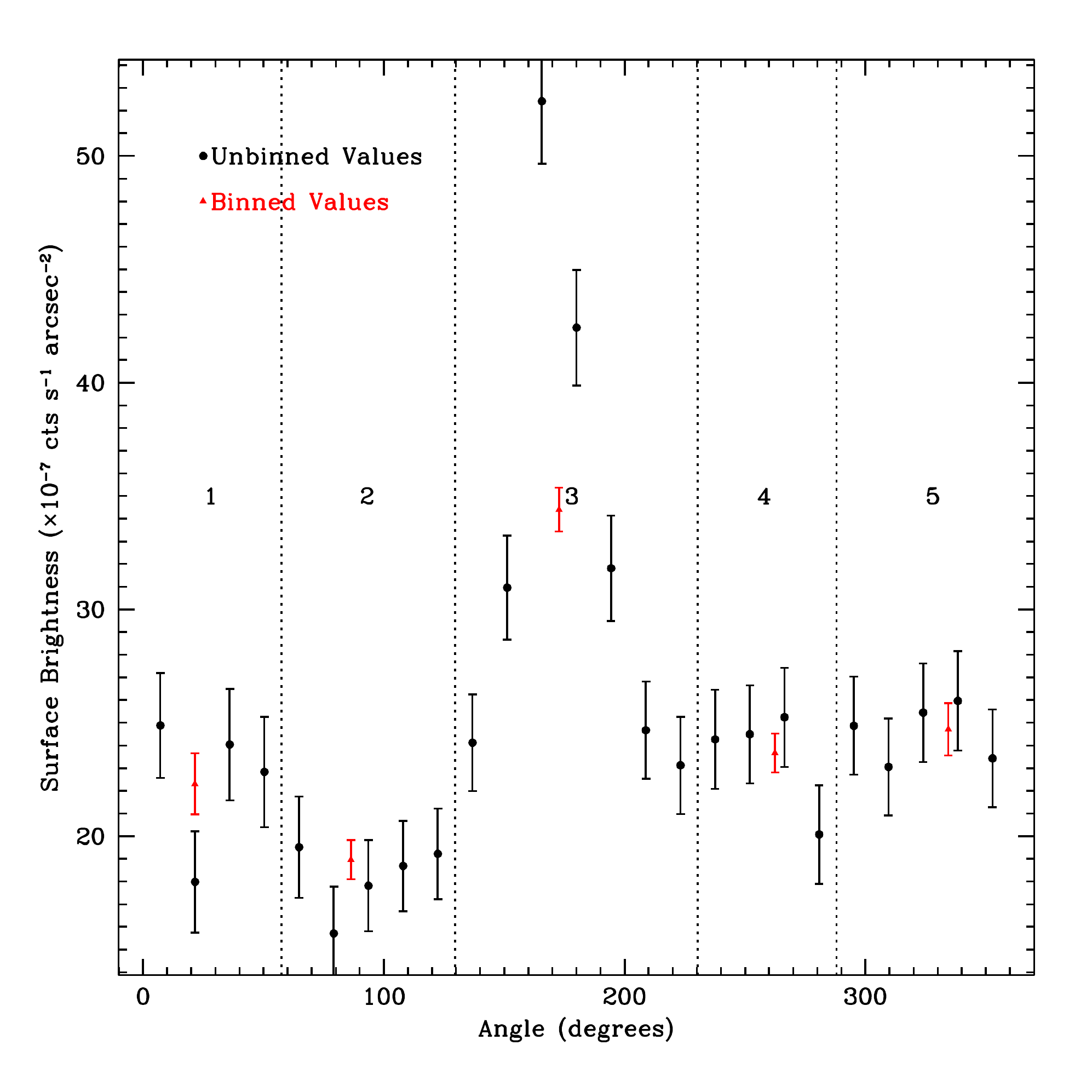}
%\vspace{-8ex}
\caption{Azimuthal surface brightness profile of A98S.  We have divided the azimuthal surface brightness profile into five sections.  The section labeled 2 contains the the northern lobe of the WAT and the section labeled 4 contains the southern lobe of the WAT.  Section 3 contains the bright, secondary subcluster core to the east of the WAT.  The surface brightness is systematically lower in the northern lobe than in the other regions of the ICM.  Sections 1 and 5 do not contain any notable features.  These sections are labeled in Figure~\ref{regions_wat}.  The binned values are shown in red triangles for reference.  The surface brightness decrement in section 4 is between $2-4\sigma$ when compared to the reference regions.}
\label{sb}
\end{center}
\end{figure}
Figure~\ref{sb} shows the azimuthal surface brightness profile.  The surface brightness of the ICM that is spatially coincident with the northern radio lobe is systematically lower than in the other regions of the ICM.  The surface brightness spikes at around $170^\circ$ in region 3.  This is due to the eastern surface brightness excess seen in Figure~\ref{a98s_beta}.  We find that the significance of the surface brightness decrement in section 2 is between $2-4\sigma$.  To determine this, we binned all of the points in each section, as labeled in Figure~\ref{sb} (each point corresponds to a wedge in Figure~\ref{regions_wat}).  We compared the surface brightness of section 2 (which contains the decrement) to the surface brightness in sections 1, 4, and 5.  The binned points can be seen in Figure~\ref{sb}.  We did not use section 3 because of the large surface brightness excess due to the secondary subcluster core.  There is no significant decrement coincident with the southern lobe.  As discussed in \S\ref{intro}, the AGN in A98 is considered a WAT radio source, despite its straight morphology, and any bending is occurring along the line of sight.  The absence of a surface brightness decrement under the southern lobe is consistent with this scenario, as the northern jet axis may be close to the plane of the sky, while the southern jet axis is not.  As a result, the southern cavity may be projected onto brighter emission at smaller physical radii, and would therefore be more difficult to detect. We also tried azimuthal binning at larger radii (150'' (290~kpc) $< r <$ 230'' (450~kpc) as compared to $r <$150" (290~kpc) for the original binning), to see if the decrement under the northern lobe was due to the structure of the subcluster, rather than the AGN.  We found that there was no significant difference in surface brightness in any of the sections.

While jets and lobes (and thus cavities in the ICM) are more commonly seen in cool core clusters, they have also been seen in clusters that have recently undergone or are currently undergoing mergers. Other examples include Cygnus A~\citep{wilson, belsole}, Abell 1446~\citep{1446}, and Abell 562~\citep{562}.

\section{Filaments and the WHIM} \label{filaments}

The three subclusters of A98 are aligned in such a way to suggest the presence of a large-scale filament.  In principle, there should be faint X-ray emission associated with this filament, so we measured the X-ray properties of the connecting region to determine if there is any evidence of the denser part of the WHIM that is expected to be found just beyond the cluster outskirts.  The redshifts of the subclusters increase from north to south, indicating that the filament is not in the plane of the sky.  The filament connecting the subcluster pairs (A98N-A98S and A98S-A98SS) is therefore tilted along the line of sight, increasing the chances of detection.  This hottest, densest part of the WHIM was observed in the cluster pair Abell~222/223 by~\citet{werner}, which has a similar orientation to A98.  Detections have also been reported in the pair Abell~399/401 using both X-ray~\citep{sakelliou} and the thermal Sunyaev-Zel'dovich (tSZ) effect~\citep{planckviii}, the pair Abell~3391/3395~\citep{tittley}.   We searched for excess emission in the regions connecting A98N-A98S and A98S-A98SS, but were unable to find any in the latter region, likely due to the short exposure time.  

\begin{figure}
\begin{center}
\plotone{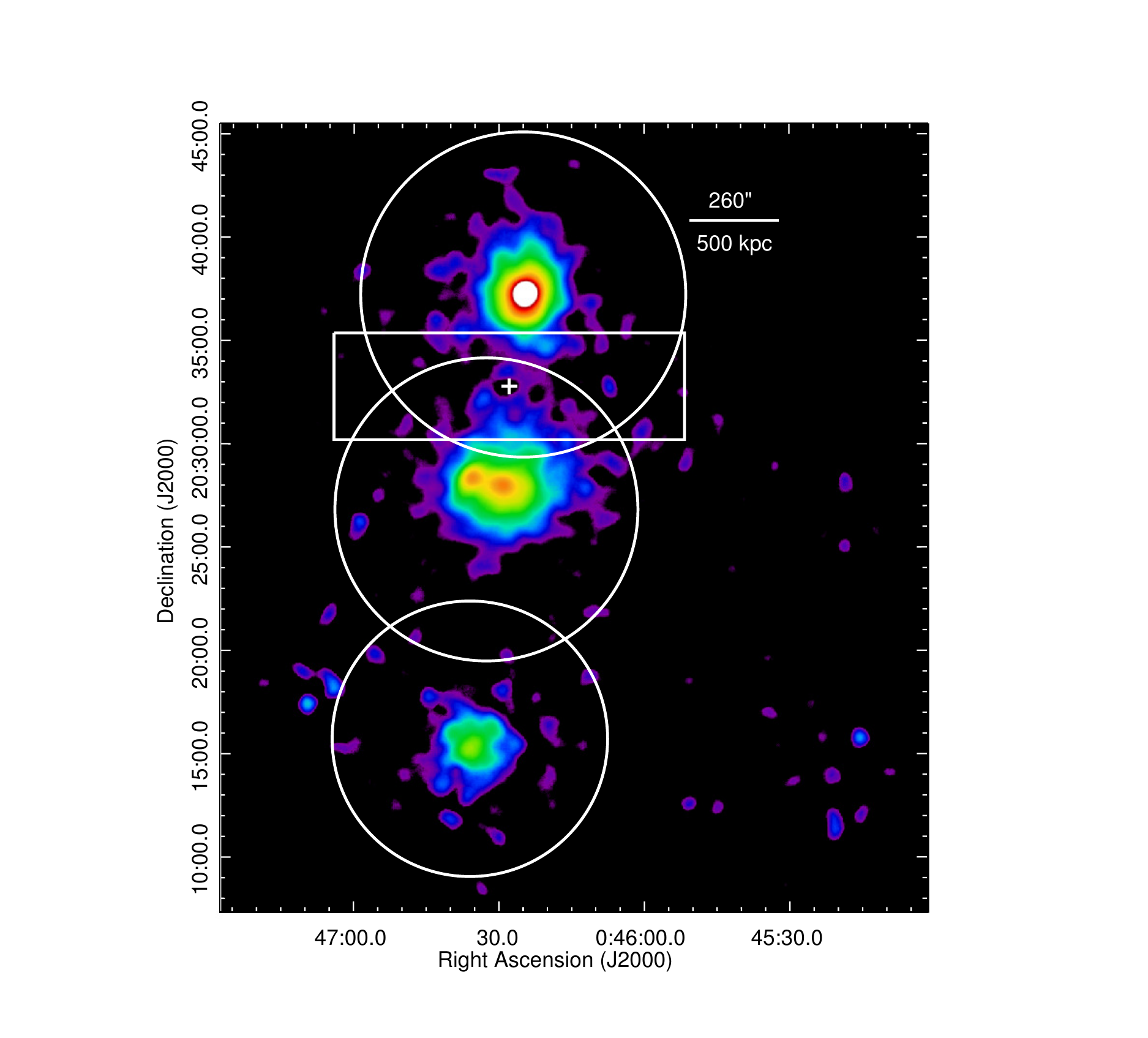}
\caption{Highly smoothed, binned \textit{Chandra} image of the diffuse emission in A98.  The image is binned by 4, and smoothed by a $40''$ Gaussian.  There is an apparent bridge of emission connecting A98N and A98S. A98N is elongated in the N-S direction.  The white box shows the region used to look for evidence of extended emission, while the white circles show the extent of $r_{500}$.  The cross marks the location of $r=0''$.}
\label{bridge}
\end{center}
\end{figure}

\begin{figure}
\begin{center}
\plotone{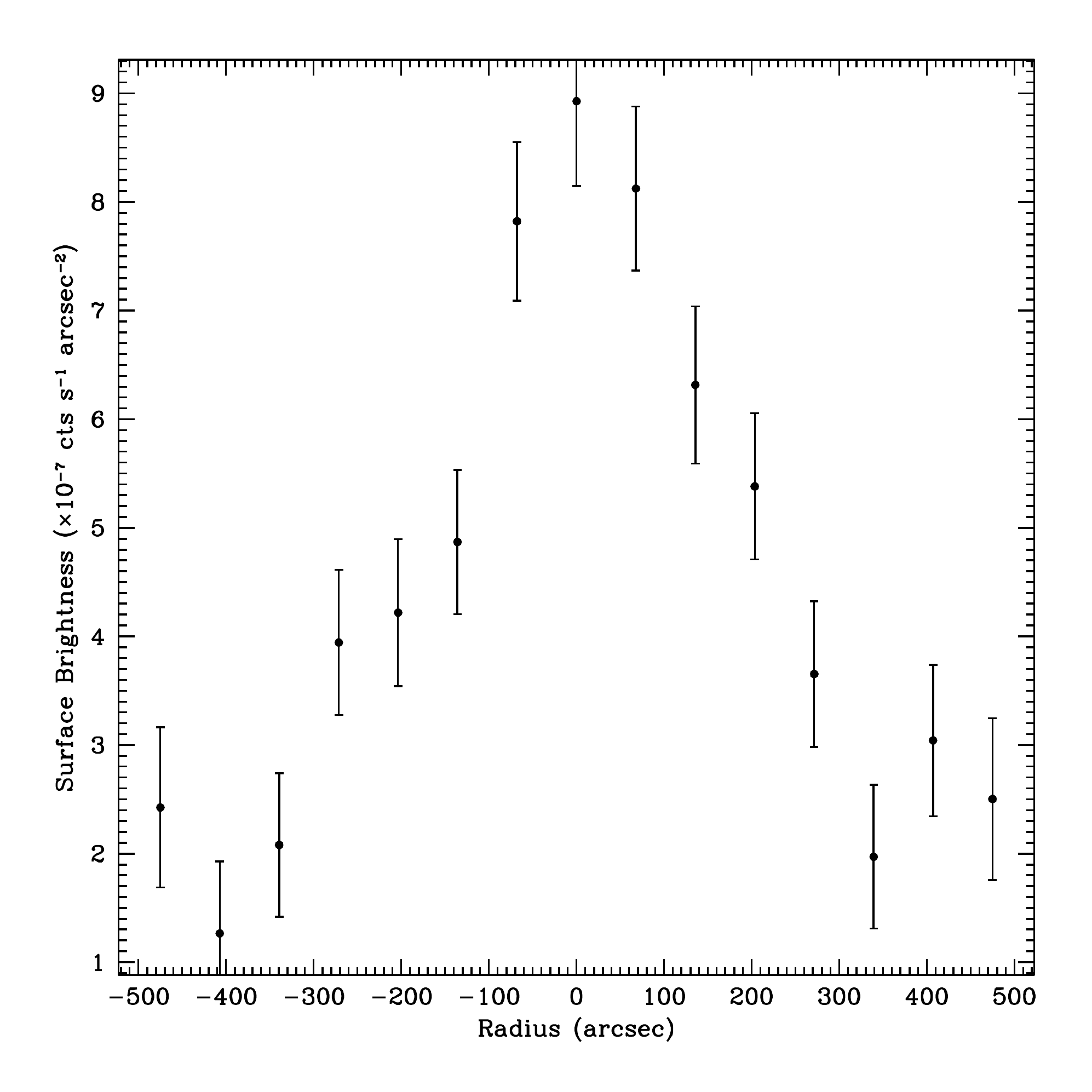}
\caption{Surface brightness profile of the region across the bridge, going E to W (shown in Figure~\ref{bridge}).  The surface brightness reaches a maximum at $r=0''$, which corresponds to the center of the bridge region.}
\label{strip}
\end{center}
\end{figure}

Figure~\ref{bridge} shows a highly smoothed, merged \textit{Chandra} image of the diffuse emission from A98. 
There is clearly emission connecting A98N and A98S.  To characterize the significance of this emission, we extracted a surface brightness profile from an E-W strip (shown in Figure~\ref{bridge}) across the bridge.  This surface brightness profile is shown in Figure~\ref{strip}.  There is a clear surface brightness peak at $r=0''$, which is the center of the bridge region.  When binned together, the three central points are significantly higher at the $6\sigma$ level when compared to points 1, 2, and 3 binned together and points 13, 14, and 15 binned together.

We then tried to fit two elliptical beta models to A98N and A98S simultaneously to model the extended emission and identify the bridge in the residual image.  However, as discussed in \S\ref{mergersig:a98n}, we could not find a model that well-described the extended emission, because it is faint and shows large-scale asymmetry.  There is also a chip gap that runs across the region of interest, giving fewer photons in the region of interest and making it difficult to fit the emission.

There are several possible explanations for the origin of the apparent bridge seen in Figure~\ref{bridge}.  The first is that it is emission from the hot, dense, inner part of the WHIM.  Second, it could be caused by the gravitational potential of the filament itself, which is distorting the gas and pulling it out from the cluster outskirts.  Third, this apparent bridge could be the result of a current or past merger, either in the form of ram pressure stripped gas or tidal interaction.  Finally, it is also possible that the feature is due to the overlap of the extended atmospheres of the two adjacent clusters.

\begin{figure}
\begin{center}
\plotone{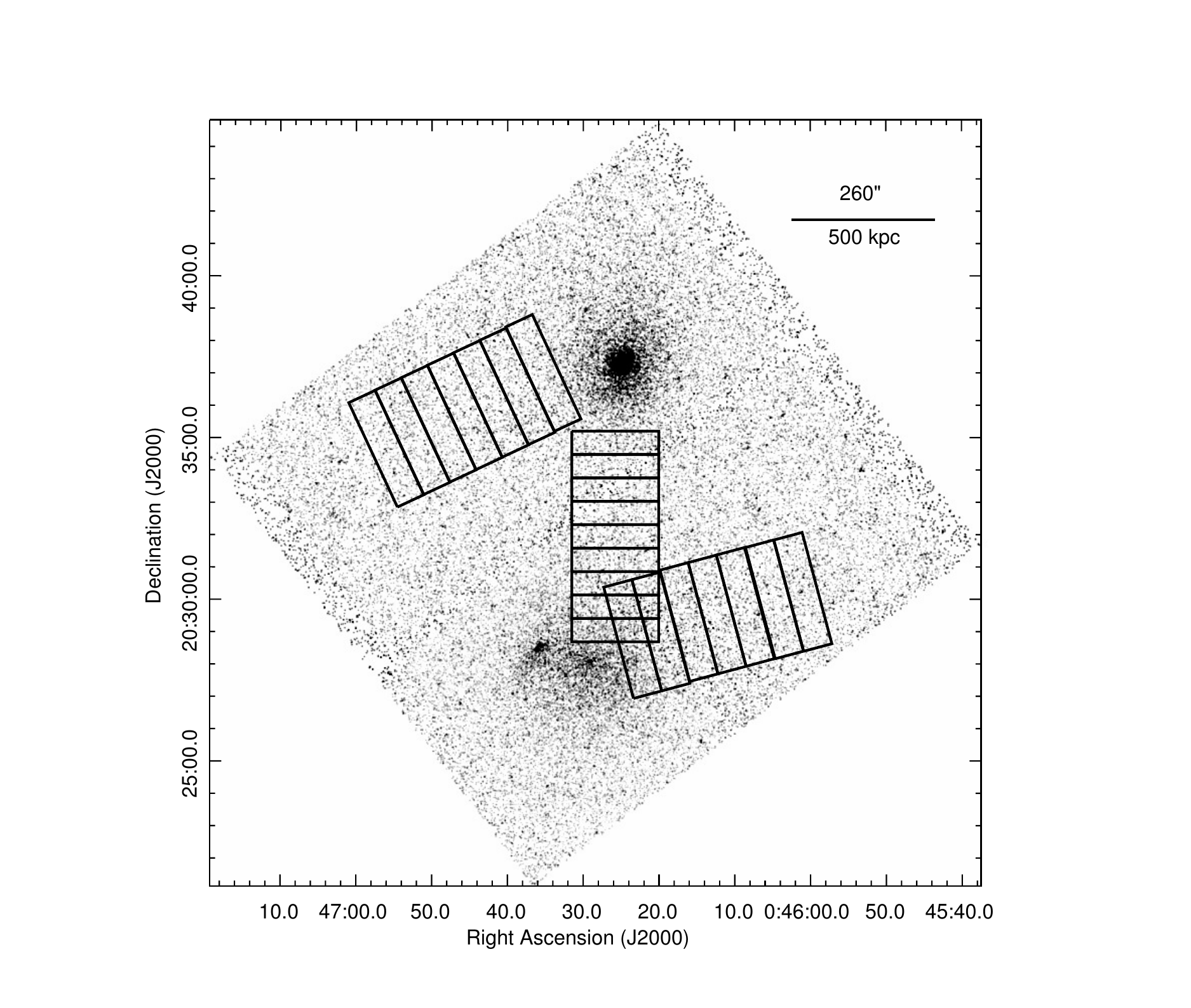}
\caption{The regions used for the surface brightness and temperature measurements.  The subdivisions were not included for the temperature measurements.  The "bridge" connects A98N and A98S, while the comparison regions extend radially outward so that we could measure the extended emission of each subcluster independently of each other.}
\label{regions}
\end{center}
\end{figure}

To distinguish among these scenarios, we compared the profile of this apparent bridge to the sum of the surface brightness profiles of the extended diffuse emission of each cluster, measured radially in regions far from the filament.  We note that even if the extended emission of each subcluster were well-described by a single beta model, we would still expect to see a surface brightness enhancement in the bridge region due to the overlapping extended atmospheres.  By choosing regions away from the potential overlap and summing them, we are able to search for X-ray emission in excess of this overlap.  The regions used for  comparison are shown in Figure~\ref{regions}.

\begin{figure}
\begin{center}
\plotone{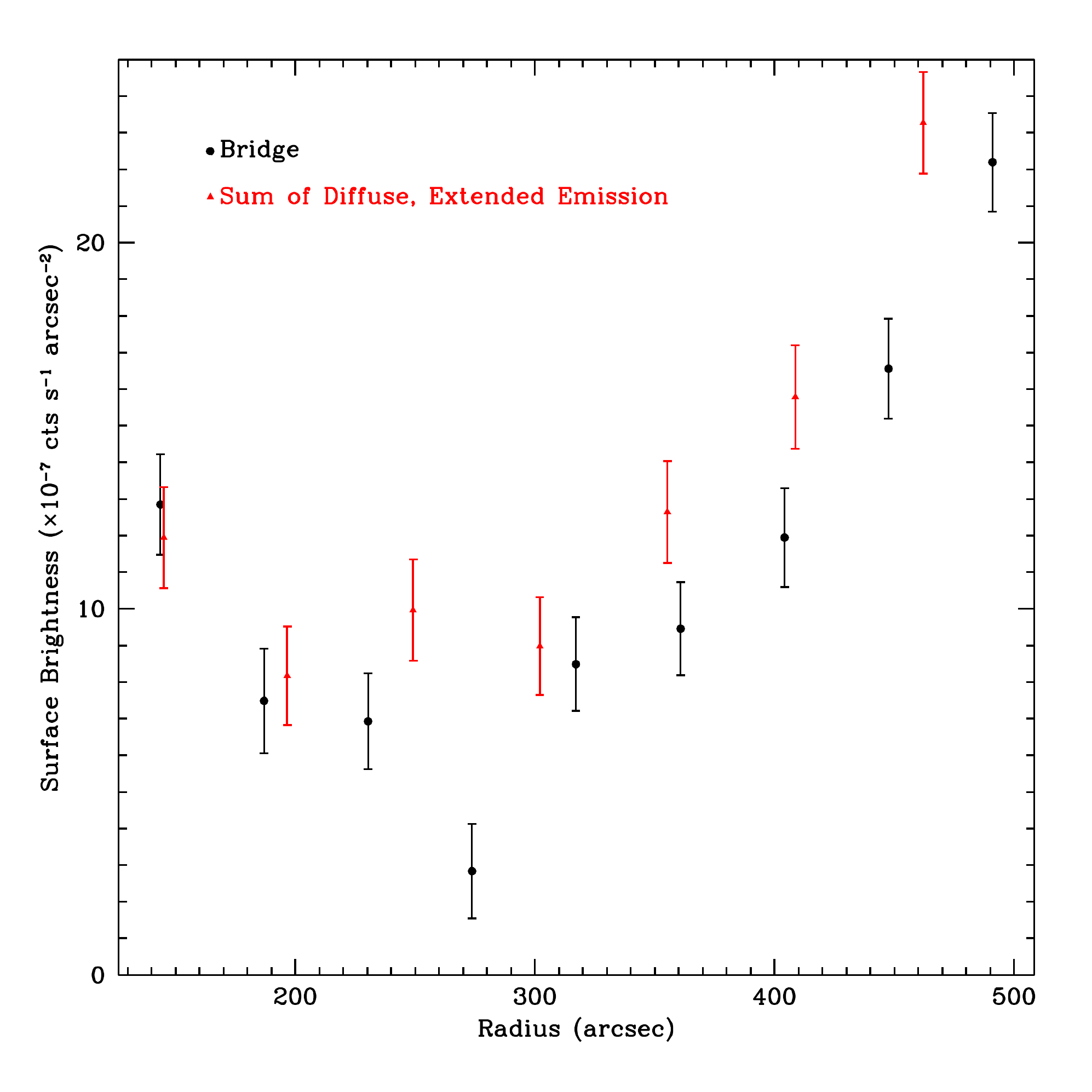}
\caption{Surface brightness profile of the bridge connecting A98N ($r=0''$) to A98S as compared to the sum of the extended diffuse cluster emission from each subcluster.  There is no significant difference between the sum of the diffuse emission and the emission from the bridge region.}
\label{whim}
\end{center}
\end{figure} 
Figure~\ref{whim} shows the comparison of the emission in the bridge region compared with the expectation from the sum of the extended emission from each subcluster. There is no significant difference between these two profiles, suggesting that the apparent bridge is consistent with arising from the overlap of the extended subcluster atmospheres.   

We also measured the temperature of the bridge region and the comparison regions using XSPEC's C-statistic due to a low number of counts. The temperature of the bridge region is $kT=2.98^{+1.11}_{-0.64}$~keV(measured with $\sim550$~counts), while the average temperature of the two reference regions is $kT=2.89^{+1.46}_{-0.66}$~keV reference region one had $\sim370$~counts and reference region two had $\sim380$~counts).  The temperature measured from the \xmm data for the bridge is $2.88^{+0.54}_{-0.38}$~keV, which is consistent with the \textit{Chandra} data (there were $1600$~counts from the MOS1 detector and 2900~counts from the MOS2 detector).   The measured temperature of the bridge region is nearly 3~keV, which is significantly hotter than the 0.01--1.0~keV range of temperatures expected for the WHIM.  For comparison, \citet{werner} find a temperature of $kT = 0.91\pm0.25$~keV for the hot, dense part of the WHIM in Abell~222/223. More counts are needed to fit multiple temperature models in this region, as we expect emission from both the relatively hot extended cluster atmospheres and possibly the dense WHIM.  

If there is emission from the hot, dense, inner part of the WHIM, then it is much fainter than the extended cluster emission, as we do not see evidence for it in Figure~\ref{whim}.  This could drive the relatively high (for the WHIM) temperature from the single-temperature fit that we found above.  While Figure~\ref{whim} shows no significant difference between the sum of the diffuse emission away from the bridge region and the bridge region, there is some evidence for tidal disruption of the extended gas halos of A98N and A98S resulting from an early stage merger between them.  In \S\ref{dynamics} we show that A98N and A98S are currently interacting.  This could explain the presence of a possible shock or spiral in A98N and the disturbance to the ICM in A98S.  Figure~\ref{bridge} shows that A98N has an N-S elongation, and that A98S is more extended to the north than the south.  These features are both consistent with tidal interaction between A98N and A98S.  Fainter emission may still be present, but better photon statistics are needed to detect faint emission and constrain multi-temperature models.
 
\section{Cluster Dynamics} \label{dynamics}

In addition to examining the subclusters in the X-ray, we searched for substructure in the optical data.  We used the individual galaxy data from \citet{pinkney} and \citet{bettoni}; however we limited our search criteria to individual galaxies within a projected distance of $\sim1$~Mpc of each subcluster center and with velocities within 1,000~km~s$^{-1}$ of the average velocity of each subcluster.  There were not enough galaxies with known redshifts that met these parameters to do a detailed optical substructure analysis.

To further constrain the dynamical state and merger history of A98, we considered whether the subclusters are bound to each other.  We followed the method outlined in \citet{beers}, although we note that there is a typographical error in their equation 13a, which is corrected in \citet{santos}.  We restricted our analysis to the A98N-A98S and A98S-A98SS pairs.  While the fact that this cluster is a triple system will in principle affect our calculations, the gravitational interactions are dominated by adjacent pairs, such that these effects are small.  There are three possibilities for each pair: bound-incoming (BI), bound-outgoing (BO), and unbound-outgoing (UO).  For the calculation, we used $t=12.1$~Gyr, which is the age of the Universe at the redshift of A98 as a whole ($z=0.1042$), rather than the redshifts of the individual subclusters.  The subcluster masses were determined using the $M-T$ relation of \citet{mt}.  We find $M_{A98N}=2.1\times10^{14}$~$M_{\sun}$, $M_{A98S}=1.8\times10^{14}$~$M_{\sun}$, and $M_{A98SS}=1.9\times10^{14}$~$M_{\sun}$.  The relative velocities ($V_r$) were found using the redshift difference between the subclusters, and the projected distance ($R_p$) was taken to be the distance between the center of each subcluster as given in \S\ref{intro}.  The redshift for A98SS was calculated using the cluster members listed in \citet{pinkney}.  Within a $7\farcm5$ ($\sim1$~Mpc) radius, there were 14 cluster members in the sample.  One of those members had a redshift of $z=0.0505$, while the BCG of A98SS has a redshift of $z=0.1205$.  This is 16.5 times the standard deviation of the other 13 cluster members.  Because of this large difference, and since there were no other cluster members with similar redshifts, we did not include this member in our calculation.  We used a total of 13 cluster members and averaged over their redshifts.  For A98N-A98S, $V_r=600$~km~s$^{-1}$ and $R_p=1058$~kpc, while for A98S-A98SS $V_r=4260$~km~s$^{-1}$ and $R_p=1428$~kpc.    

For the case where the clusters are gravitationally bound, the equations of motion can be written in the following parametric form~\citep{beers, santos}:
\begin{equation}
R=\frac{R_m}{2}\left(1 - \cos\chi\right),
\label{bound1}
\end{equation}

\begin{equation}
t=\left(\frac{R^3_m}{8GM}\right)^{1/2}\left(\chi - \sin\chi\right),
\label{bound2}
\end{equation}

\begin{equation}
V=\left(\frac{2GM}{R_m}\right)^{1/2}\frac{\sin\chi}{\left(1-\cos\chi\right)},
\label{bound3}
\end{equation}
where $R_m$ is the separation of the subclusters at maximum expansion, $M$ is the total mass of the system, $t$ is the age of the Universe at the redshift of the cluster, and $\chi$ is the development angle used to parameterize the equations.  For the case where the clusters are not gravitationally bound, the parametric equations are:
\begin{equation}
R=\frac{GM}{V^2_\infty}\left(\cosh\chi-1\right),
\label{unbound1}
\end{equation}

\begin{equation}
t=\frac{GM}{V^3_\infty}\left(\sinh\chi-1\right),
\label{unbound2}
\end{equation}

\begin{equation}
V=V_\infty\frac{\sinh\chi}{\left(\cosh\chi-1\right)},
\label{unbound3}
\end{equation}
where $V_\infty$ is the asymptotic expansion velocity.  $V_r$ and $R_p$ are related to the system parameters by
\begin{equation}
V_r=V\sin\alpha,~R_p = R\cos\alpha.
\label{params}
\end{equation}
In these equations, $\alpha$ is the projection angle of the system with respect to the plane of the sky.  Finally, energy considerations determine the limits of the bound solutions:
\begin{equation}
V_{\rm r}^2  R_{\rm p} \leq 2GM~{\rm sin^2 ~ \alpha ~cos ~\alpha}.
\label{energy}
\end{equation}

\begin{figure}
\begin{center}
\plotone{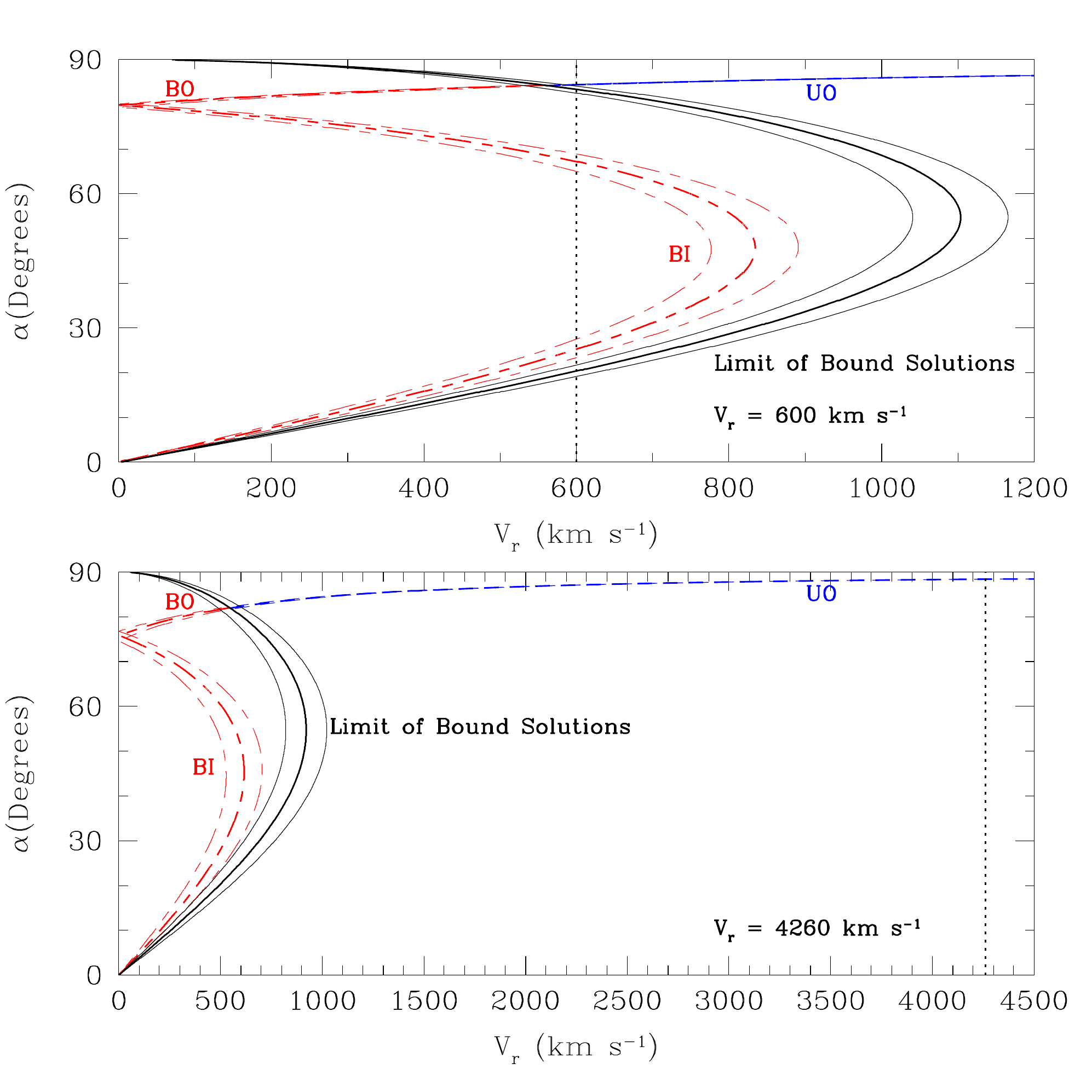}
%\vspace{-8ex}
\caption{The top panel shows the dynamical solutions between A98N and A98S, while the bottom panel shows the solutions between A98S and A98SS.  The red, dashed-dotted lines show the bound solutions, while the blue, dashed lines show the unbound solutions.  The black, solid lines show the limit of the bound solutions.  For each of the solutions, the outer lines show the solutions based on the 90\% errors on the temperature measurements.  The vertical dotted line is the relative velocity between the subclusters, which is also labeled in each panel.  There are no bound-outgoing solutions between A98N and A98S, and there are no bound solutions at all between A98S and A98SS.}
\label{dynamical}
\end{center}
\end{figure}

Figure~\ref{dynamical} shows the different solutions for each of the subcluster pairs, as well as the solutions based on the 90\% error bars on the temperature measurements.  There are no BO solutions for A98N-A98S, and no bound solutions at all for A98S-A98SS.  For  A98N-A98S, there are two BI solutions due to the ambiguity in the projection angle, $\alpha$: a) $\chi=4.4$~rad, $\alpha=67.1^\circ$, $R=2.72$~Mpc, $R_m=4.16$~Mpc, $V=651$~km~s$^{-1}$; b) $\chi=5.1$~rad, $\alpha=25.2^\circ$, $R=1.17$~Mpc, $R_m=3.83$~Mpc, $V=1411$~km~s$^{-1}$.  The unbound solution is: $\chi=0.8$~rad, $\alpha=84.4^\circ$, $R=10.8$~Mpc, $V=603$~km~s$^{-1}$, $V_\infty=234$~km~s$^{-1}$.  The three-dimensional separations between A98N and A98S are 2.7~Mpc, 1.2~Mpc, and 10.8~Mpc, for the three solutions, respectively.  For A98S-A98SS, the unbound solution is:  $\chi=7.11$~rad, $\alpha=88.5^\circ$, $R=53.2$~Mpc, $V=4262$~km~s$^{-1}$, $V_\infty=4254$~km~s$^{-1}$.  The three-dimensional separation between A98S and A98SS is 54.5~Mpc.   We used the method discussed in \citet{girardi} to determine the probability that the system is bound.  The probability is computed using the solid angle $P=\int_{\alpha_1}^{\alpha_2} \cos\alpha\mathrm{d}\alpha$, where $\alpha_1$ and $\alpha_2$ are the angles at which the limit of bound solutions intersect $V_r$.  We find that there is a 67\% probability that A98N and A98S are bound.  With no BO solution, it is unlikely that A98N and A98S have encountered each other in the past.

Our solution for the A98N-A98S pair differs from the solution found by \citet{beers}.  They find that the system is bound with a 98\% probability.  They also find that there are only bound solutions -- two incoming and one outgoing.  The two BI solutions they find are: a) $\chi=4.06$~rad, $\alpha=78^\circ$, $R=3.67$~Mpc, $R_m=4.57$~Mpc, $V=550$~km~s$^{-1}$; b) $\chi=5.37$~rad, $\alpha=13^{\circ}$, $R=0.76$~Mpc, $R_m=3.90$~Mpc, $V=2452$~km~s$^{-1}$.  Their BO solution is: $\chi=1.58$~rad, $\alpha=86^\circ$, $R=9.51$~Mpc, $R_m=18.8$~Mpc, $V=541$~km~s$^{-1}$.  Rather than use X-ray observations, they used optical data to determine the mass of the system, and found that it was $6.6\times10^{14}$~$M_{\sun}$.  They also used optical data to determine the average radial velocity of each subcluster and found $V_r=539$~km~s$^{-1}$.  They used $R_p=740$~kpc and an age of the Universe at $z=0$ of 9.8~Gyr (not the age of the Universe at the redshift of the cluster), which was determined using a $q_0=0$, $H_0=100$~Mpc~km$^{-1}$~s$^{-1}$ cosmology.  In contrast, we used a cosmology where $H_0=70$~km~s$^{-1}$, $\Omega_{\Lambda}=0.7$, and $\Omega_M=0.3$.  For the A98N-A98S pair, we used $R_p=1058$~kpc and $V_r=600$~km~s$^{-1}$.  To determine the probability of each solution, \citet{beers} integrated to determine the fractional area of the sphere corresponding to each solution.  Since the radial velocity regions for each solution are not equally probable, they assumed a Gaussian distribution of velocities centered around $V_r=539$~km~s$^{-1}$, integrated, and multiplied by the fractional areas.

\section{Summary} \label{summ}

We have presented both \textit{Chandra} and \xmm observations of the triple cluster system A98.  The global spectrum of each subcluster is well-described by a single APEC model with the $N_H$ fixed to the Galactic value determined by \citet{nh}.  We find that A98N has kT$\sim3.1$~keV, with an abundance of $Z\sim0.40$~$Z_\sun$, A98S has a temperature of kT$\sim2.8$~keV and an abundance of $Z\sim0.20$~$Z_\sun$, and A98SS has a temperature of kT$\sim3.0$~keV and an abundance of $Z\sim0.45$~$Z_\sun$.  The results from the \textit{Chandra} and \xmm data are consistent with each other. 

A98N shows a clear asymmetry in the surface brightness map (Figure~\ref{a98n_beta}), and a possible asymmetry in the temperature distribution (Figure~\ref{tmap}).   For this feature to be a sloshing spiral, the brighter region is expected to be cooler than the surrounding gas, and A98N and A98S would have to have been perturbed by a merger at least 0.3~Gyr ago~\citep{am, johnson}.  X-ray spectral analysis shows that the surface brightness excess has a warmer temperature at the $1.4\sigma$ level compared to gas at the same radius to the north and compared to the gas to the south of the excess, where there is no surface brightness excess.  While this is not highly significant, combined with the dynamical history of the cluster (which suggests that this is the first interaction of A98N and A98S) this feature is consistent with a shock forming due to the merger between A98N and A98S.  The location of the surface brightness excess (south of A98N, along the merger axis) also suggests that if this is a shock, it is forming because of the merger.  The measured temperatures in the region of the asymmetry favor the idea that this is a merger bow shock, with a Mach number of $M=1.3$.  The dynamical history discussed in \S\ref{dynamics} also favors the merger shock scenario, as there is no indication that there is a bound-outgoing orbit.  Because there is no BO orbit, it is unlikely that these two subclusters have undergone a previous encounter.  If they had, they would most likely be bound to each other based on the overdensity collapse model of structure formation.  

A98S is host to a WAT AGN, which is coincident with the warmer part of the subcluster.  The northern lobe of this AGN is evacuating a cavity in the ICM, causing a significant surface brightness decrement.  Although there are a few examples of cavities in non-cool core clusters, they are relatively rare. 

Figure~\ref{a98} shows that both A98S and A98SS have a disturbed morphology.  The temperature map (Figure~\ref{tmap}) shows that A98S also has a dual temperature structure, which is confirmed via spectroscopic measurements of the two distinct temperature regions.  The optical data for both of these clusters reveals the presence of two dominant galaxies, consistent with the interpretation that these subclusters are experiencing ongoing or late stage mergers.  

An analysis of the cluster dynamics shows that A98N and A98S have a 67\% chance of being bound, while A98SS is not bound to A98S.  Although A98S is dominated by a later-stage merger happening in the east-west direction, there is also some elongation to the north, which is also shown in Figure~\ref{bridge}.  The north-south elongation seen in both subclusters could be due to tidal interaction occurring during the merger.  While the X-ray image shows that there is an apparent bridge of emission between the two subclusters, a detailed surface brightness profile shows that this bridge is consistent with overlapping extended emission from A98N and A98S.  This region is where we would expect to see the signature of the hottest, densest part of the WHIM.  The measured temperature of this region is high for the WHIM, and consistent with the temperature of the gas at similar radii outside of the bridge region.  A deeper observation is required to possibly separate the WHIM from the observed emission, as well as apply a two-temperature model to the X-ray gas in the region of interest.  The dynamical history is very similar to the triple system PLKG214.6+37.0, which was originally detected using the tSZ effect, and then observed with \xmm~\citep{planckvi}.  In that system, an apparent bridge is also detected between two subclusters, which is consistent with the overlapping extended emission from said subclusters.  As in A98, the third component of this system is likely unbound.  A98S and A98SS have a disturbed morphology, indicating that they have both undergone recent mergers.  The merger in A98SS is separate from the merger between A98N and A98S.

\section*{Acknowledgements}
RPM was partially supported by ADP grant NNX13AE83G and partially supported by the National Science Foundation through NSF AST-1313229.  SWR was supported by the Chandra X-ray Center through NASA contract NAS8-03060, the Smithsonian Institution, and by {\it Chandra} X-ray observatory grant GO2-13161X.

This research has made use of the NASA/IPAC Extragalactic Database (NED) which is operated by the Jet Propulsion Laboratory, California Institute of Technology, under contract with the National Aeronautics and Space Administration.

Funding for SDSS-III has been provided by the Alfred P. Sloan Foundation, the Participating Institutions, the National Science Foundation, and the U.S. Department of Energy Office of Science. The SDSS-III web site is http://www.sdss3.org/.

SDSS-III is managed by the Astrophysical Research Consortium for the Participating Institutions of the SDSS-III Collaboration including the University of Arizona, the Brazilian Participation Group, Brookhaven National Laboratory, Carnegie Mellon University, University of Florida, the French Participation Group, the German Participation Group, Harvard University, the Instituto de Astrofisica de Canarias, the Michigan State/Notre Dame/JINA Participation Group, Johns Hopkins University, Lawrence Berkeley National Laboratory, Max Planck Institute for Astrophysics, Max Planck Institute for Extraterrestrial Physics, New Mexico State University, New York University, Ohio State University, Pennsylvania State University, University of Portsmouth, Princeton University, the Spanish Participation Group, University of Tokyo, University of Utah, Vanderbilt University, University of Virginia, University of Washington, and Yale University.

\bibliographystyle{apj}
\bibliography{a98}

\end{document}